# Evaluation of medium range machine learning models for sub-seasonal prediction

**Catherine de Burgh-Day[1*], Chen Li[1], Debra Hudson[1], Li Shi[1], Harrison Cook[2^], Robin Wedd[1], Griffith Young[1]**

[1]Australian Bureau of Meteorology 700 Collins Street, Docklands VIC 3008, Australia

[2]European Centre for Medium-Range Weather Forecasts (ECMWF)

[^]Employed by the Bureau of Meteorology at the time this work was done

**Corresponding Author:** catherine.deburgh-day@bom.gov.au

## Abstract

The performance of two machine learning (ML) atmosphere models - GraphCast and FourCastNetV2 - is evaluated in the context of sub-seasonal prediction, including their ability to represent key climate drivers of variability, namely the Madden-Julian Oscillation and the Southern Annular Mode. Model skill is assessed over both a 38-year hindcast period and a 2.5-year hindcast period. The longer period overlaps with the training windows of the ML models but provides a larger sample for robust evaluation, while the shorter period is independent of the ML model training period. This dual evaluation illustrates a compromise approach to the problem of insufficient independent data for evaluation of the models for sub-seasonal prediction. The ML models are compared against the Bureau of Meteorology's physics-based seasonal prediction system, ACCESS-S2, for the 38-year period, and a more recent physics-based coupled model for the shorter hindcast period. Across the two evaluation periods, both ML models have surprisingly good skill for sub-seasonal timescales, given they were designed for forecasting on medium range timescales. In general, the ML models are as skilful as the physical model ensemble mean at shorter lead times and comparable to the physical model ensemble members at longer lead times.

# 1 Introduction

The science of sub-seasonal prediction has grown rapidly over the last decade (e.g., Robertson and Vitart 2019, Domeisen et al. 2022) and forecasts from operational centres are now commonplace. Forecasts of weather and climate on sub-seasonal timescales are highly beneficial to a variety of stakeholders, and are increasingly being used for decision-making (e.g., White et al. 2022). The Australian Bureau of Meteorology has been providing sub-seasonal forecasts to the public since August 2019 using ACCESS-S (Australian Community Climate and Earth System Simulator–Seasonal; Hudson et al., 2017; Wedd et al., 2022).

These operational sub-seasonal forecasts currently use traditional physics-based models. However, the field of data-driven weather and climate forecasting is advancing rapidly. Several operational centres are now developing, evaluating and operationalising machine learning (ML) models and exploring their strengths and weaknesses compared to physics-based models. There are still several key questions and active research around aspects such as the physical consistency of ML-generated forecasts (e.g., Bonavita, 2023; Hakim & Masanam, 2024), appropriate ensemble generation (e.g., Lang et al. 2024b; Price et al. 2025; Liu et al., 2025), stability for forecasts beyond ~14-days (e.g., Galluser et al. 2025), skill and use in an operational-like context (e.g., Ben Bouallègue et al., 2024; DeMaria et al., 2025; Harrison et al., 2025), skill across a range of meteorological phenomena and for predictions of extremes (e.g. Charlton-Perez et al. 2024; Sun et al, 2025; Sahu et al, 2025; Morisseau et al. 2025), and fairness in ML model development and training (Olivetti & Messori, 2025).

Most of the focus of these efforts has been on short- to medium-range forecasts (0-10 days) (e.g., Pathak et al., 2022; Lam et al, 2023; Bi et al., 2023; Chen et al., 2023; Lang et al., 2024a; Alet et al., 2025; Dunstan et al., 2025; Xiong et al., 2025). However, the research and development on ML models for sub-seasonal and seasonal prediction is growing (e.g., Chen et al. 2024; Lang et al., 2024b; Kent et al. 2025; Bonev et al., 2025; Antonio et al. 2025). One of the major benefits of ML models is that they are orders of magnitude computationally cheaper and faster to run than our traditional models. Using ML models for sub-seasonal and seasonal prediction would significantly reduce the large computational burden and time taken to produce hindcasts. It also opens other possibilities, including higher resolution forecasts, the production of larger ensembles and more frequent system upgrades.

In this paper, the performance of two machine learning models - GraphCast (Lam et al., 2023) and FourCastNetV2 (Bonev et al., 2023, July) - is evaluated for sub-seasonal climate prediction, including a focus on key climate drivers such as the Madden-Julian Oscillation (MJO) and the Southern Annular Mode (SAM), which significantly influence sub-seasonal variability over Australia.

These two ML models are trained on 6-hourly data representing the three-dimensional state of the atmosphere, derived from reanalysis and analysis datasets such as ERA5 (Hersbach et al., 2020). Through this training, the models learn how the atmospheric state evolves over time and how different variables are related to each other. When run in forecast mode, they predict the atmospheric state 6 hours ahead, then use that prediction as input to forecast the next 6 hours, and so on. This auto-regressive approach means the model relies on its own previous outputs to generate future predictions. Note that this method does not necessarily guarantee a realistic or stable climate, as the focus is on short-term atmospheric evolution rather than capturing broader climate patterns or the conservation of key quantities. It is also fundamentally different from predicting a climate index like Niño3, which targets a specific large-scale index of variability rather than the full atmospheric state.

Because ML weather and climate models are trained on a large proportion of the high-quality historical reanalysis period, a limited sample remains for independent evaluation. For example, GraphCast was trained on data from 1979 to 2021 (Lam et al., 2023), leaving only a few years of recent data for independent evaluation. While this is sufficient for evaluating weather phenomena, it becomes limiting when analysing longer timescale processes, and when evaluating climate drivers or extremes. As such, two evaluation periods are used in this study. The first involves benchmarking GraphCast and FourCastNetV2 against the Bureau of Meteorology's seasonal prediction system, ACCESS-S2 (Wedd et al, 2022), using a 38-year hindcast period from 1981 to 2018. Although this period provides a substantial dataset for analysis, it overlaps with the training data of both ML models. Therefore, a second, independent evaluation period from January 2022 to June 2024 is introduced, which falls outside the training windows. For this shorter period, we compare the ML models to forecasts from the UK Met Office's more recent global coupled model configuration, GC5 (Global Coupled Model version 5) (note that ACCESS-S2 uses an older coupled model, GC2). GC5 is being tested at the Bureau over this recent period. GC5 was not used for the longer evaluation due to the impracticality of generating a comparable hindcast dataset given resource constraints.

By evaluating GraphCast and FourCastNetV2 across both periods, we obtain a robust sample from the 38-year hindcast and an independent evaluation from the 2.5-year period. Consistent performance across these two windows enhances confidence in our results. However, it is important to acknowledge that there are several limitations of this approach. Since the 38-year evaluation period overlaps with the training data, it likely leads to an optimistic estimate of the ML model skill. Meanwhile, the shorter 2.5-year period, while independent, includes a very limited number of relevant events, such as MJO occurrences and SAM phase transitions, and does not support the calculation of anomalies relative to its own climatology (due to insufficient samples to compute a robust climatology). As such, this compromise should not be viewed as a substitute for a larger, fully independent evaluation dataset in future work.

Our choice of ML models is motivated by their different architectures and each of them being likely to be well suited to the task of sub-seasonal prediction. GraphCast has a graph-based architecture and is popular for medium range forecasts because it shows higher skill than many competitors (Rasp et al., 2024), which may extend into longer range forecasts. FourCastNetV2 uses a Spherical Fourier Neural Operator architecture and is popular for applications aiming for long-term stability and lower growth of biases (e.g., Watt-Meyer et al., 2023; Guan et al., 2024). It is important to emphasise, however, that neither of these models is designed for sub-seasonal prediction and would not typically be considered suitable for applications at these timescales without further adaptation. This assessment therefore represents a preliminary investigation into the potential of ML atmospheric models for sub-seasonal prediction. Another limitation of the current study is the absence of ensemble forecasts for the ML models, which is essential for operational applications of sub-seasonal prediction. The generation of ensembles by perturbing initial conditions for ML models designed as single member systems, such as GraphCast and FourCastNetV2, has yielded mixed results (Weyn et al., 2021; Mahesh et al., 2024; Brenowitz et al., 2025; Liu et a., 2025). As such, it is preferable to generate ensembles using ML models designed specifically for generating ensembles (e.g. Lang et al., 2024; Price et al., 2025), which were not openly available when this study was initiated. Given these challenges, this study should be viewed as a contribution to initial explorations into the potential of machine learning-based atmospheric models for sub-seasonal prediction, alongside related work such as Diao & Barnes (2025). It aims to inform and guide future research by highlighting key limitations, methodological considerations, and opportunities for advancing the development and evaluation of such models.

The remainder of this paper is structured as follows: In Section 2, the data and methods are described; in Section 3 the results of the evaluation across the 38-year and 2.5-year periods are presented; in Section 4 a discussion of these results is provided; and conclusions are given in Section 5.

# 2 Data and methods

## 2.1 Reanalysis Data

The ML and physical models are evaluated against the ERA5 reanalysis dataset (Hersbach et al., 2020). ERA5 is based on the ECMWF Integrated Forecasting System, has a horizontal resolution of ~31 km, 137 vertical levels, and hourly output. Since the ML models are trained on ERA5, it is valuable to also assess their performance against alternative reanalysis datasets. We therefore conducted an additional evaluation using the NCEP/NCAR reanalysis for all MJO and SAM related analyses, which yielded results consistent with those obtained from ERA5. For brevity, we present only the results against ERA5.

The reanalysis datasets were mapped to the resolutions of ACCESS-S2 and GC5 (both described below) using bilinear interpolation for the evaluations over the 38-year and 2.5-year periods, respectively.

## 2.2 The ACCESS-S2 model

ACCESS-S2 (Australian Community Climate and Earth-System Simulator - Seasonal, version 2) has been the Bureau's seasonal prediction system since October 2021 (Wedd et al., 2022). ACCESS-S2 uses the Met Office global coupled model version 2 (GC2) at N216 resolution (~60km) and 85 levels in the atmosphere, 0.25° resolution and 75 vertical levels in the ocean, four land surface layers and five sea-ice thickness categories. Details of the data assimilation and ensemble generation can be found in Wedd et al (2022). The ACCESS-S2 data used in this study are 9-member ensembles from the hindcast set, with ensemble member 0 being the unperturbed central member.

## 2.3 Global coupled model version 5 (GC5)

Global Coupled model version 5 (GC5) is the latest GC model release from the Met Office. It is comprised of the Global Atmosphere and Land configuration GAL9.0 - based on the Unified Model (UM) version 12.2 and the Joint UK Land Environment Simulator (JULES) version 6.3; and the Global Ocean and Sea Ice configuration GOSI9.0, which includes the Nucleus for European Modelling of the Ocean (NEMO) version 4.0 and the Sea Ice modelling Integrated Initiative ($SI^3$) version 4.0 (Xavier et al. in preparation).

GC5 was run with an N320 resolution (~40km at the equator) for the atmosphere-land and 0.25 degrees for the ocean-ice. A 9-member ensemble was used for the experiments, using pre-perturbed Met Office Global Ensemble Prediction System (MOGREPS) (Bowler et al, 2008) initial conditions obtained from the Met Office. The central member has no stochastic physics perturbations, while the remaining members have these perturbations turned on for the forecast integration.

## 2.4 Machine learning models

GraphCast (Lam et al., 2023) is a deep learning model for the atmosphere, developed by Google Deepmind, which uses a graph-based neural network architecture. Three configurations of the model have been trained and made publicly available (https://github.com/google-deepmind/graphcast). For this work, the "GraphCast_operational" version is used. This version has 0.25-degree resolution, a 6-hour timestep, 13 pressure levels in the atmosphere, and is trained on ERA5 reanalysis (Hersbach et al., 2020) data from 1979 to 2017 before being fine-tuned on ECMWF IFS-HRES data from 2016 to 2021.

FourCastNetV2 (Bonev et al., 2023, July) is a successor to FourCastNet (Pathak et al., 2022) and is also a deep learning model for the atmosphere. FourCastNetV2 innovated on the original by moving from a Fourier Neural Operator (FNO) model with a Vision Transformer (ViT) backbone to a Spherical Fourier Neural Operator (SFNO)-based model. This involved reformulating the FNO architecture to a spherical geometry (i.e., SFNO), and removing the ViT backbone to substitute patch embedding with down-sampling in the initial SFNO layer of the model to save memory. Unlike FNOs, SFNO-based architectures respect spherical symmetry by construction, leading to improved stability when simulating the atmosphere or other natively-spherical systems. FourCastNetV2 is trained on ERA5 data from 1979 to 2015 with a timestep of 6-hours, 0.25-degree resolution and 13 pressure levels in the atmosphere.

In this study, both GraphCast and FourCastNetV2 forecasts were initialised with ERA5 reanalysis data, and produced outputs on the ERA5 grid. Like the reanalysis data, these outputs were then mapped to the ACCESS-S2 and GC5 model grids using bilinear interpolation for the evaluations.

### 2.5 Experiment design and methods

As described in the Introduction, two evaluation periods were used in this study – a longer 38-year period which overlaps with the GraphCast and FourCastNetV2 training windows, and a smaller 2.5-year period which is independent. The ML model forecasts were run with an autoregressive timestep of 6 hours, starting from 00Z on the 1st and 16th of each month in the 2.5-year hindcast, and from 00Z on the 1st of each month in the 38-year hindcast.

The GC5 model was run over the 2.5-year hindcast period only, with the same lead time and start date-time configuration as the ML models. ACCESS-S2 hindcast data were used for comparison with the ML models from the 1st of each month of the full 38 year hindcast. A summary of the details of the hindcasts used in this analysis are presented in Table 1.

## 3 Results

### 3.1 The 38-year hindcast period

A detailed evaluation was undertaken using the 38-year hindcast period for FourCastNetV2, GraphCast, and ACCESS-S2. Consistent with standard practice in sub-seasonal prediction (e.g., Hudson et al. 2017; Wedd et al. 2022), the evaluation is focussed on the skill of predicted anomalies. Evaluation of anomalies enables assessment of skill beyond variations in climatology. Anomalies are calculated relative to the lead-time dependent hindcast climatology of each model respectively. For the ACCESS-S2 ensemble members, the ensemble mean climatology is used.

Table 1: Summary of the two hindcast periods

| Name | Period | Starts (per month) | Total number of forecast cases | Models |
|---|---|---|---|---|
| 2.5-year hindcast | Jan 2022 – Jun 2024 | 00Z 1st and 16th | 60 | • GraphCast<br>• FourCastNetV2<br>• GC5 9-member ensemble |
| 38-year hindcast | Jan 1981 – Dec 2018 | 00Z 1st | 456 | • GraphCast<br>• FourCastNetV2<br>• ACCESS-S2 9-member ensemble (taken from existing hindcast set) |

#### *3.1.1 Weekly skill*

Globally-averaged weekly mean correlation skill of 10m u-wind speed, 850 hPa temperature, mean sea level pressure, and rainfall anomalies for weeks 1 to 4 are presented in Figure 1. Correlations are calculated using the latitude-weighted Pearson's correlation coefficient. Note that rainfall is not included as a variable in FourCastNetV2, therefore all rainfall evaluation is limited to GraphCast.

For all variables in week 1, the ML models have slightly higher skill than ACCESS-S2. This is not surprising since these models have been designed for weather forecasting, whereas the focus of ACCESS-S2 is for sub-seasonal and seasonal prediction. For weeks 3 and 4 the ML models show similar skill to that of a single member of ACCESS-S2. In contrast, the ACCESS-S2 ensemble mean has the highest skill in weeks 3 and 4. This raises the question of how competitive an appropriately designed ML model ensemble could be, and to what extent ensemble design and choice of perturbation scheme would influence its performance. This is demonstrated by Lang et al. (2024), who adopt an ensemble-based training method with a Continuous Ranked Probability Score (CRPS) loss function for their ensemble ML model, AIFS-CRPS, which outperforms the uncalibrated physical model ensemble.

Maps of weekly correlation skills of screen-level temperature anomalies for weeks 1 and 3 are shown in Figure 2. The map for ACCESS-S2 shows the correlation skill of a single ensemble member rather than the ensemble mean. The spatial pattern of skill in week 1 is broadly consistent across the models and aligns with expectations (e.g., Zhu et al 2014). Skill is highest in the extratropics, due to the greater predictability of baroclinic systems, and lowest in the tropics except for a region of high skill in the equatorial eastern Pacific which is linked to the El Niño Southern Oscillation (ENSO). In general, both

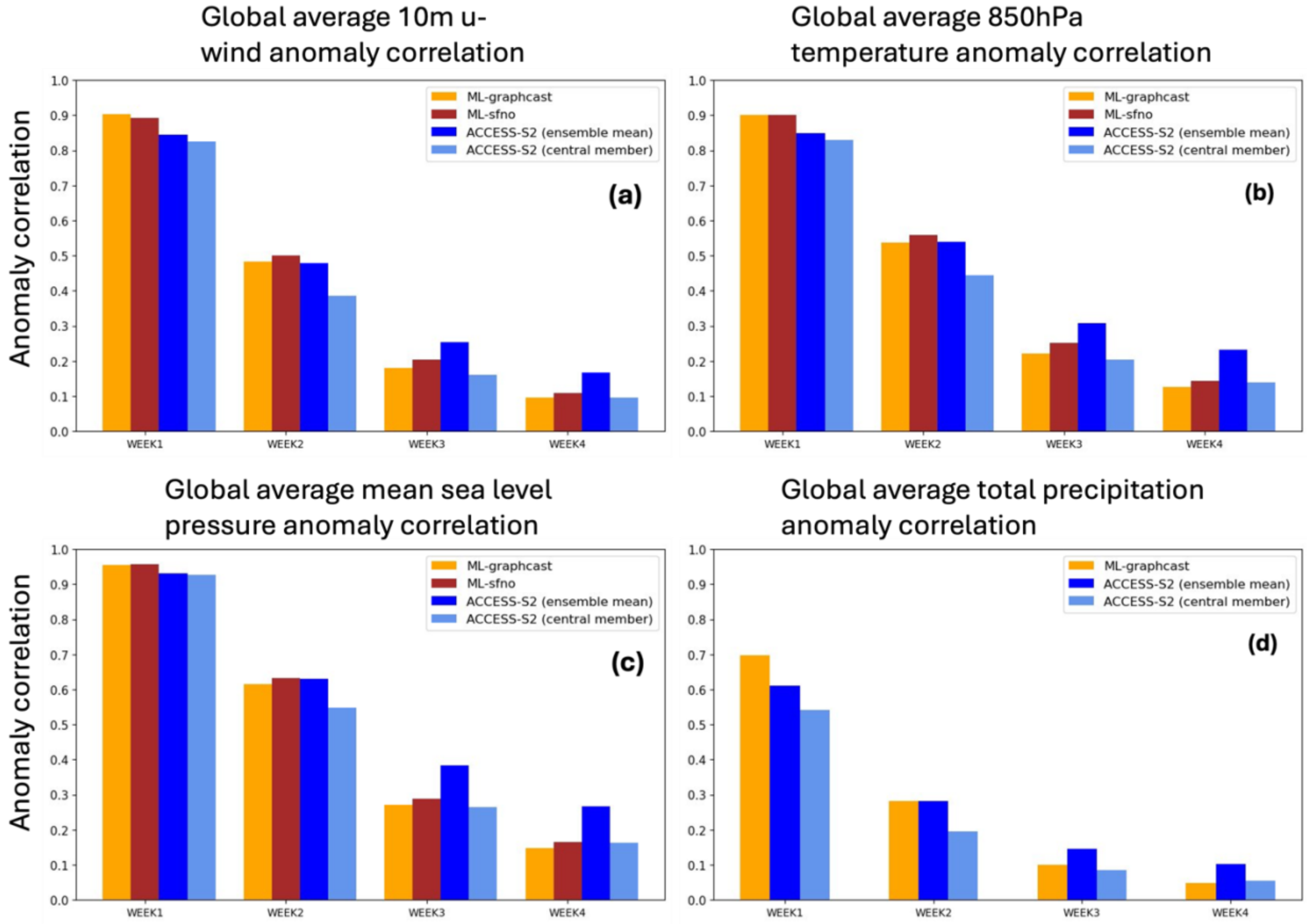

Figure 1: Globally averaged correlation skills of (a) 10m u-wind speed, (b) 850 hPa temperature, (c) mean sea level pressure, and (d) rainfall for lead weeks 1 to 4 for the 38-year hindcast. Yellow, red, dark blue and light blue bars show the correlations for GraphCast, FourCastNetV2, the 9-member ACCESS-S2 ensemble mean and a single member of ACCESS-S2 respectively. These correlation skills are computed based on weekly anomalies relative to each model's own climatology for the period 1981–2018, considering all forecast start dates within this time period.

ML models show higher skill at week 1 than ACCESS-S2, particularly over sub-tropical land and high latitudes. Interestingly, FourCastNetV2 is significantly more skillful than ACCESS-S2 and GraphCast over the equatorial western Pacific and Indian Ocean regions, including the Maritime Continent. In week 3 extratropical skill has dropped compared to week 1 and most of the predictability lies in the tropics and over the oceans. The ML models show similar patterns and regions of skill to ACCESS-S2, but overall perform less well, including in the high skill region of the eastern equatorial Pacific. At this longer lead time, the inclusion of an ocean model in ACCESS-S is probably contributing to the skill through the representation of lower-frequency ocean and coupled processes. In contrast, while the ML models use surface temperature as a predictor, they do not include any sub-surface ocean information.

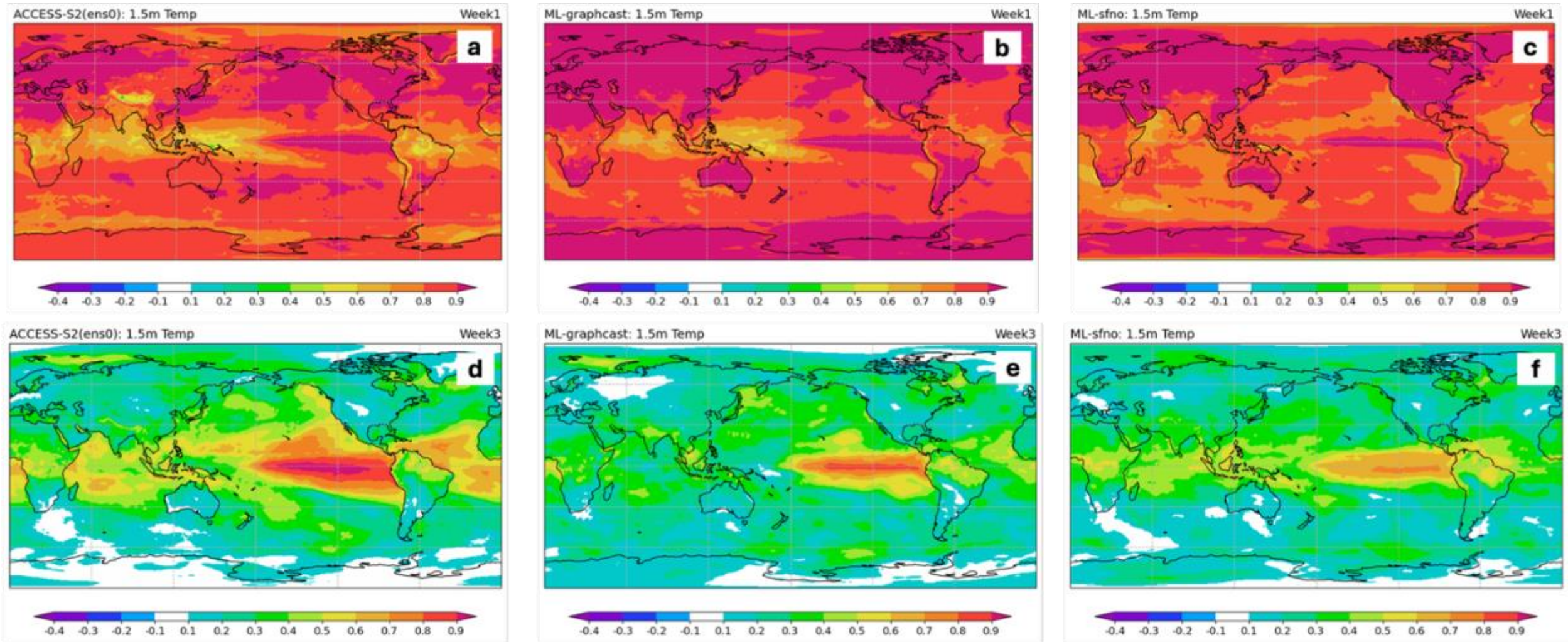

Figure 2: The upper-row figures illustrate the anomaly correlation skills of weekly temperature at 1.5m at lead time week 1 across the 38-year hindcast period for: a single ACCESS-S2 ensemble member (a), GraphCast (b), and FourCastNetV2 (c). The bottom-row figures (d–f) represent the same variables as the upper row but at lead time week 3. These correlation skills are computed based on weekly anomalies relative to each model's own climatology for the period 1981–2018, considering all forecast start dates within this time period.

Similar results are found for rainfall, in that GraphCast out-performs ACCESS-S2 in week 1 but is not as skilful in week 3 (Figure 3). Encouragingly, the regions of skill for GraphCast generally align well with those of ACCESS-S2.

### *3.1.2 The MJO*

The MJO is an eastward-propagating mode of tropical tropospheric variability characterised by large scale convective and circulation anomalies typically recurring every 30-60 days. It is the dominant mode of tropical variability and predictability on sub-seasonal timescales, with impacts extending away from the tropics including over Australia, where rainfall anomalies are closely linked to the MJO's phase (e.g. Wheeler et al., 2009; Marshall & Hendon, 2019). While the MJO has an estimated intrinsic predictability of around 5-7 weeks, the current generation of physical models are able to skilfully predict the MJO up to 20-45 days in advance depending on season and phase (e.g., Ding & Seo, 2010; Neena et al. 2014; Jiang et al. 2020).

The MJO is typically measured using the bivariate Real-time Multivariate MJO (RMM) index (Wheeler and Hendon 2004). The RMM index is calculated from a combined empirical orthogonal function (EOF) analysis of 15°S–15°N-averaged outgoing longwave radiation (OLR) and 850- and 200-hPa zonal wind anomalies. However, since neither GraphCast nor FourCastNetV2 produce OLR, a wind-only version of the index is used here, whereby only two variables (850- and 200-hPa u-wind) are projected onto the standard observation-based EOFs instead of three. This approach has been used in other studies examining the MJO in ML models which do not provide OLR (e.g., Lang et al., 2024).

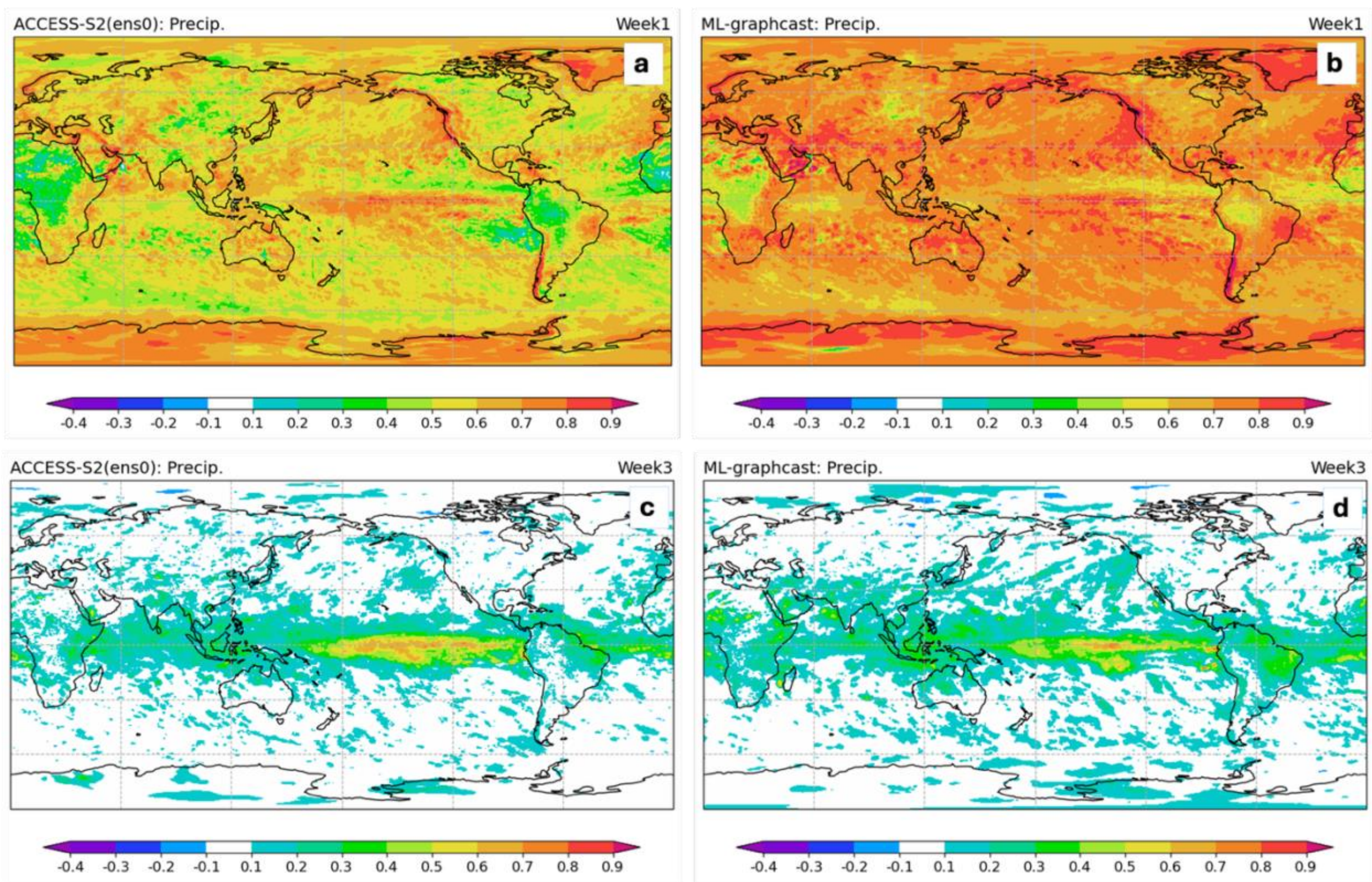

Figure 3: As in Figure 2 except for precipitation.

Two MJO cases in the 38-year hindcast period are presented in Figure 4, with initialisation dates of the 1st March 2012 and the 1st February 2017, and prediction lead-times of 28 days. The vertical black line on each Hovmöller marks 150˚ E, indicating the eastern edge of the Maritime Continent region. For the 2012 case, the MJO phase diagram and the 850 hPa u-wind anomaly Hovmöllers show that the MJO was in phase 3 (i.e., eastern Indian Ocean) at the start of the forecast and remained active as it propagated across the Maritime Continent, eventually weakening over the western Pacific. ACCESS-S2 (both the ensemble mean and the central member) and FourCastNetV2 represented the magnitude and speed of the event well to approximately lead day 17, after which they failed to capture the strengthening of the event seen in ERA5. In contrast, GraphCast struggled to capture the event, propagating slowly and becoming inactive over the Maritime Continent. The 2017 event started in phase 4 (i.e., over the Maritime Continent) and then intensified, remaining active throughout the entire four-week period. In general, all the models under-estimated the amplitude of the MJO over the first two weeks of the forecast. In this case, in contrast to the 2012 case, GraphCast performed relatively better than the other models. Analysis of these two MJO cases suggests that the ML models can effectively reproduce MJO structure and propagation. Their outputs resemble those of ACCESS-S2 (or any other physical model) such that, without prior knowledge, one might easily mistake them for traditional physical models.

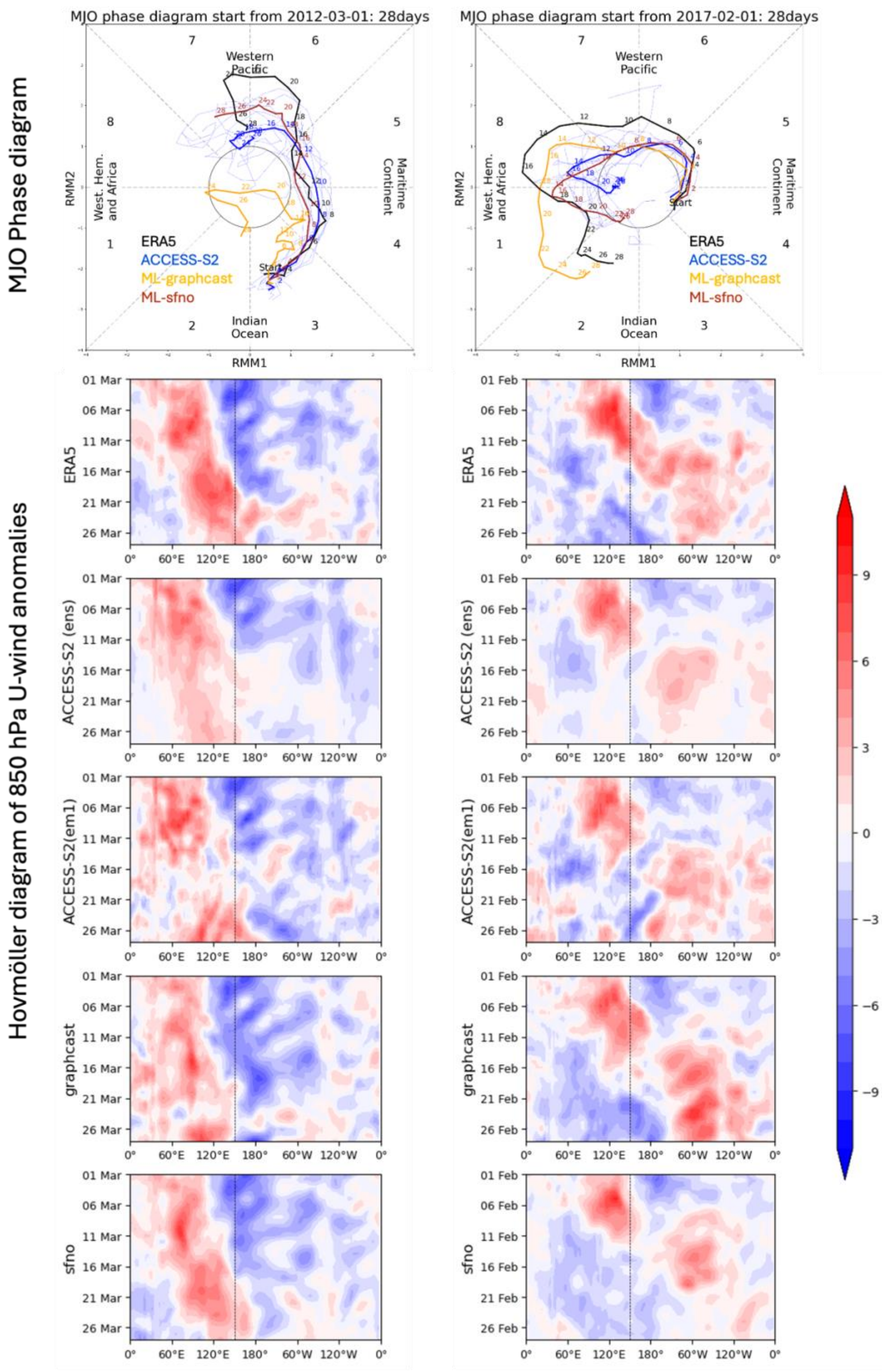

Figure 4: The phase diagram (top row) for two MJO events, beginning from the 1st of March 2012 (left column) and the 1st of February 2017 (right column), and the corresponding U850 anomalies evolution for each event in (from top to bottom) ERA5, the ACCESS-S2 ensemble mean, one ACCESS-S2 ensemble member, GraphCast, and FourCastNetV2.

MJO skill across all forecast start-dates in the 38-year period are shown in Figure 5. Skill is calculated as the bivariate correlation of the RMM1 and RMM2 indices, and the root-mean-squared-error (RMSE) of the magnitude of the combined RMM index, respectively. At shorter lead-times (<15 days) FourCastNetV2 has the highest skill, after which the most skilful model is the ACCESS-S2 ensemble mean. FourCastNetV2 is, however, more skilful than individual ACCESS-S2 ensemble members in terms of correlation and RMSE out to 30-days lead-time. GraphCast has similar correlation and RMSE performance to the ACCESS-S2 ensemble mean in the first two weeks. At longer lead times, out to 250-days, it is still competitive with individual ACCESS-S2 members in terms of correlation, but has a higher RMSE. Figure 6 shows the average MJO amplitude with lead time, where it can be seen that Graphcast significantly overestimates the MJO signal beyond 2 weeks, while ACCESS-S2 and FourCastNetV2 are more similar and both slightly underestimate it relative to ERA5. Interestingly, GraphCast shows larger U-850 hPa wind biases over the tropical Indian Ocean and Maritime Continent region than FourCastNetV2 and ACCESS-S2 after the first week of the forecast (Figure 7), which is probably related to the somewhat poorer MJO skill and overestimated MJO magnitude in GraphCast at longer leads.

To evaluate the ability of each model to capture the magnitude and phase speed of MJO events as they move over the Maritime Continent - which is often a challenge for physical models - composite phase diagrams were constructed for all events in the 38 year hindcast period that were active for more than 14 days and that started in phases 1, 2 and 3 respectively (Figure 8 left column). To complement these, plots of the mean composite phase location as a function of lead-time were produced (Figure 8 right column), providing a measure of the propagation speed of the MJO for a given starting phase, relative to that in ERA5. If a model has a composite phase propagation line that is steeper than that of ERA5, then the MJO is propagating too quickly, whereas if it is shallower, the MJO is propagating too slowly. In general, ACCESS-S2 and FourCastNetV2 perform better in capturing MJO amplitude compared to GraphCast, with their composite trajectories closely matching the reanalysis. In contrast, certain individual MJO events from GraphCast produced anomalously large amplitudes that exceeded the ERA5 envelope, contributing to the elevated RMSE and amplitude observed in Figures 5 and 6. Additionally, the composite MJO propagation speed in GraphCast was systematically slower across all initial phases 1-3 after about 14 days and deteriorated further with increasing lead-time. ACCESS-S2 showed good composite MJO propagation speed in phases 1 and 3, but propagated too slowly in phase 2 from around

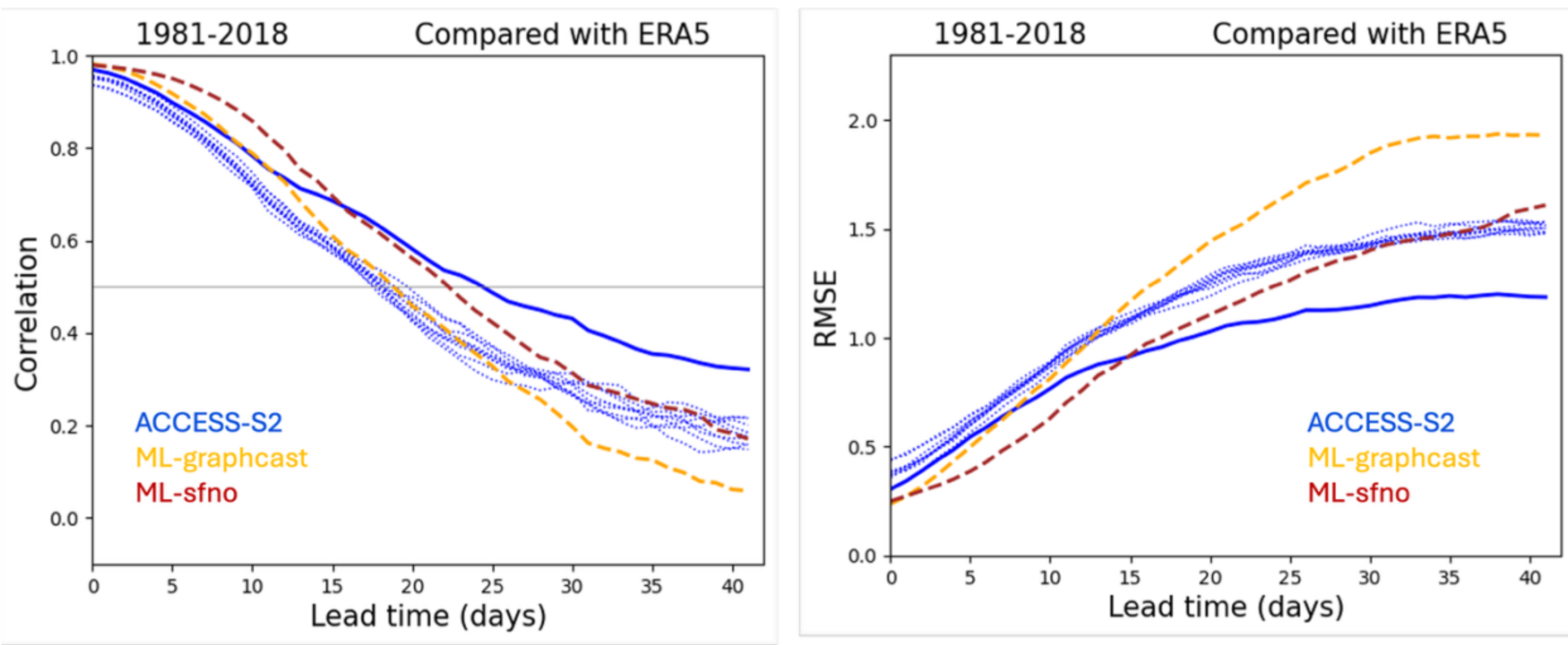

Figure 5: MJO anomaly correlation (left) and RMSE (right) skill for all months in all years from 1981 to 2018 for ACCESS-S2 ensemble members (dotted blue), the ACCESS-S2 ensemble mean (solid blue), GraphCast (yellow) and FourCastNetV2 (red).

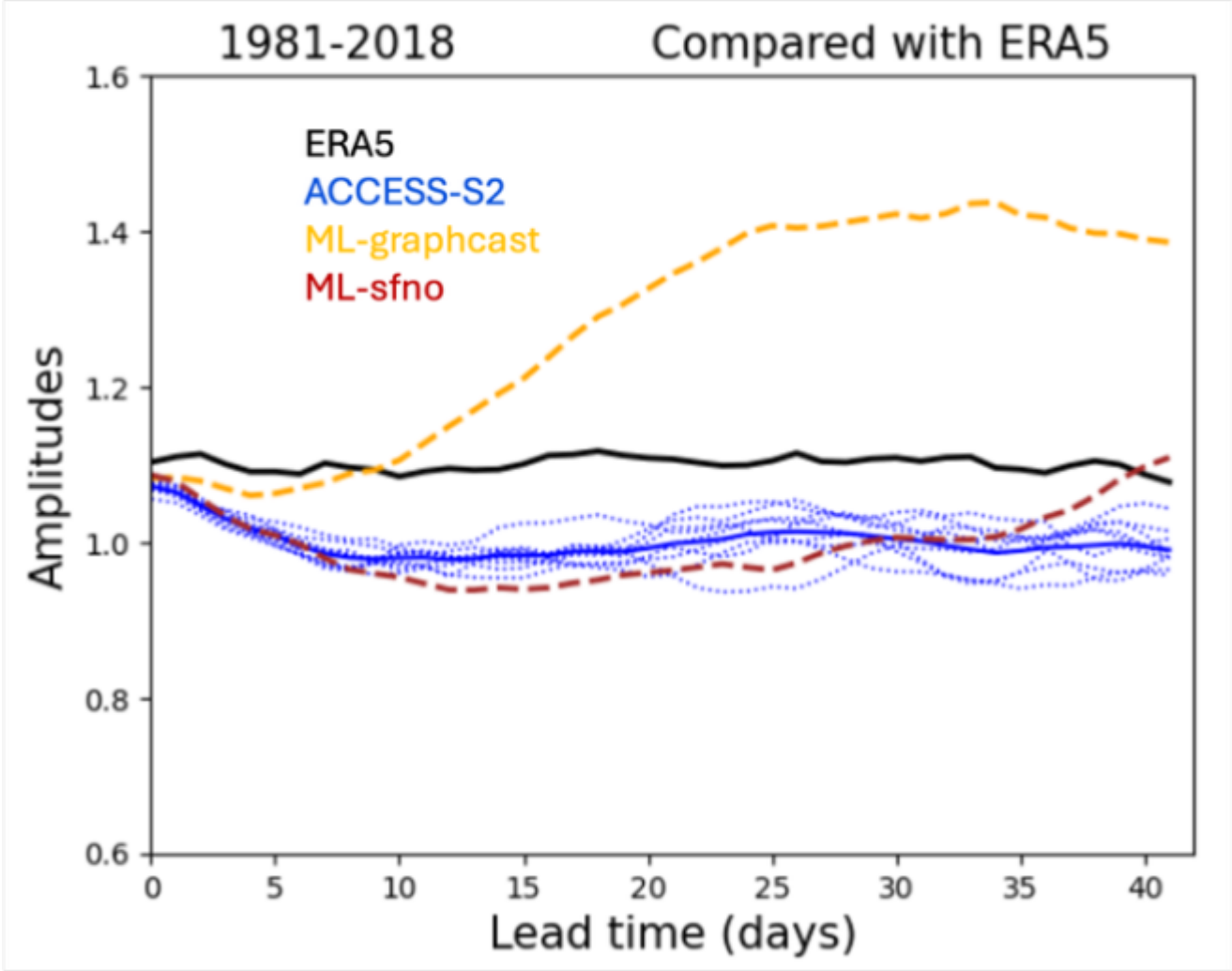

Figure 6: Average MJO amplitude with lead time for all months in all years from 1981 to 2018 for ACCESS-S2 ensemble members (dotted blue), GraphCast (dashed yellow) and FourCastNetV2 (dashed red). The solid blue line shows the average of the amplitudes of the ACCESS-S2 ensemble members, not the amplitude of the ACCESS-S2 ensemble mean. The black line is the average MJO amplitude for ERA5 for all months in all years from 1981 to 2018, matched to each forecast lead time.

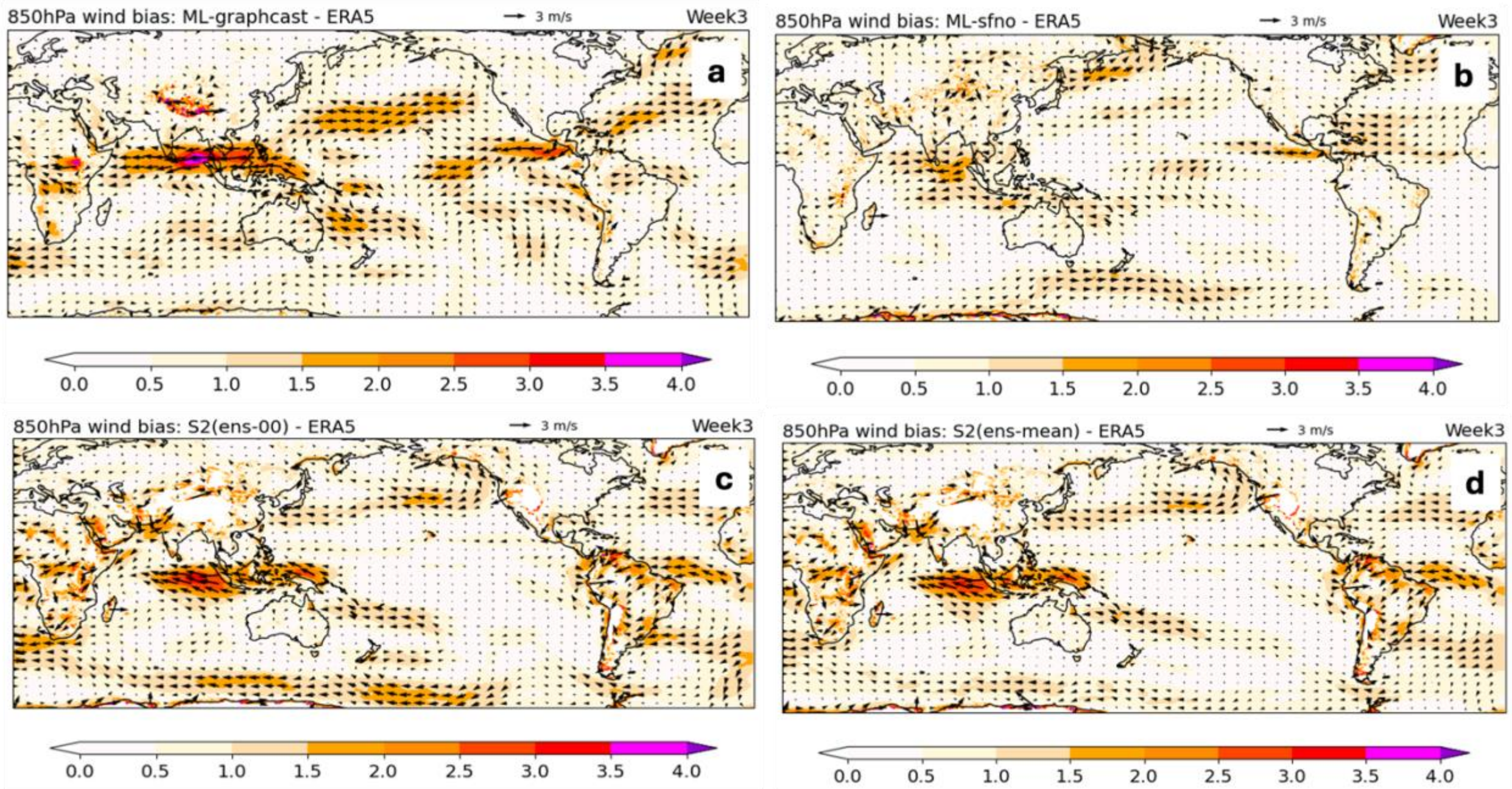

Figure 7: Weekly-average 850 hPa annual mean wind biases relative to ERA5 across the 38-year hindcast period for (a) GraphCast, (b) FourCastNetV2, (c) one ACCESS-S2 ensemble member, and (d) the ACCESS-S2 ensemble mean, at lead week 3. The biases are calculated by computing weekly means for each lead time, followed by annual means across model runs, and then subtracting from the equivalent from ERA5, before averaging the difference across the 38-year hindcast period.

15 days. FourCastNetV2 matched the phase speed of ERA5 best in phases 2 and 3 for all lead times, and also did so phase 1 to about 20 days, after which it was too slow (Figure 8 right column). Interestingly, while FourCastNetV2 had the best overall performance in terms of composite phase speed, it tended to underpredict the number of MJO events, with fewer than the number identified in ERA5 in the 38-year hindcast period for all three phases examined. In contrast, GraphCast had a tendency to overpredict the number of events in all phases, while ACCESS-S2 had too few events in phase 1, and too many in phase 3.

Finally, the MJO plays a critical role in modulating rainfall across the tropical Indo-Pacific, and the ability of a model to capture the associated rainfall patterns is essential for skilful sub-seasonal rainfall prediction. Figure 9 shows the composites of daily rainfall anomalies associated with each MJO phase. For ERA5, the composites are based on all days in November to March (NDJFM) across the 38-year period. For the models, they represent the relationship during the first (top row) and third weeks (bottom row) of the forecast, using NDJFM initialisation dates from the hindcast. These composites reflect the coincident relationship between the model's forecasted MJO phase on a given day and the corresponding

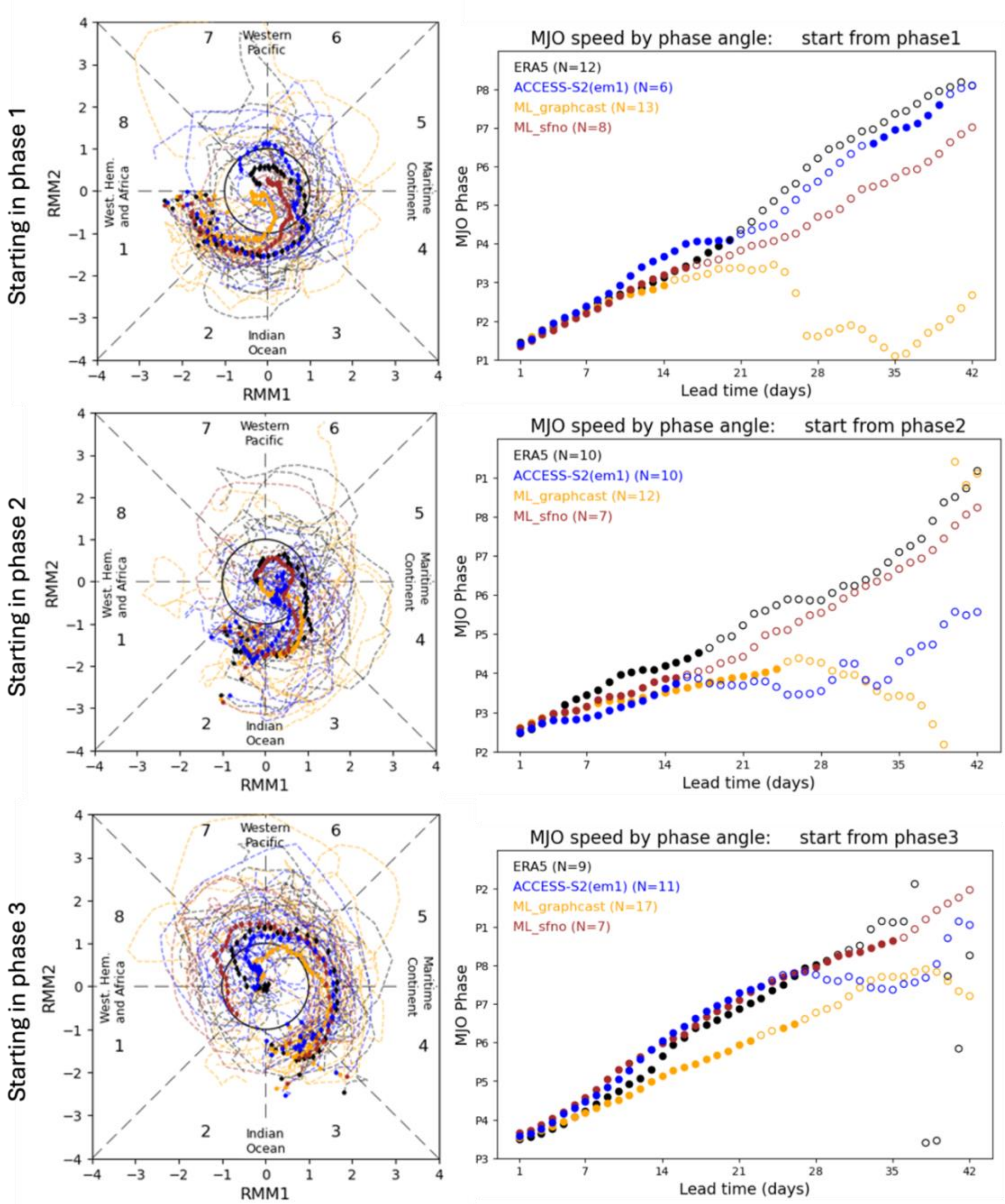

Figure 8: Left column: Phase diagrams for all MJO events in the 38-year hindcast period active for more than 14 days starting in phase 1 (top), phase 2 (middle) and phase 3 (bottom) for ACCESS-S2 (blue), GraphCast (yellow), FourCastNetV2 (red), and ERA5 (black). The dashed lines show individual MJO events, and the solid lines show the composite of all events. Right column: Composite phase speed plots for all MJO events in the 38-year hindcast period active for more than 14 days starting in phase 1 (top), phase 2 (middle) and phase 3 (bottom) for ACCESS-S2 (blue), GraphCast (yellow), FourCastNetV2 (red), and ERA5 (black). On this plot, a faster propagating MJO event has a steeper slope, and a slower event has a shallower slope. Open circles indicate a composite amplitude that is weak

(less than 1) while solid circles indicate a composite amplitude greater than 1. The number of cases contributing to the figure for each dataset is indicated in the legend of the right column plots.

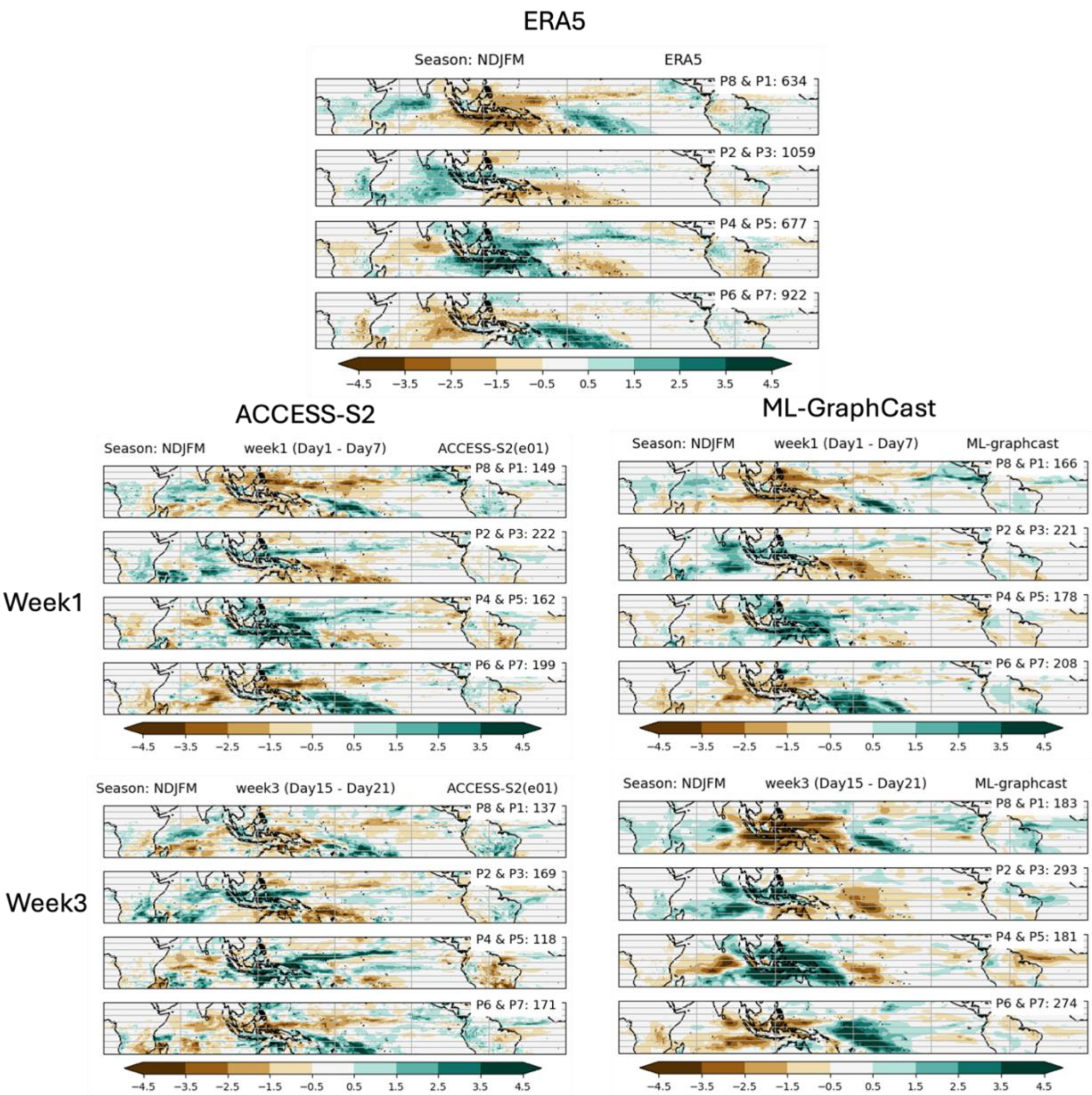

Figure 9: Rainfall composite maps for MJO events in NDJFM across the 38-year hindcast period, for all MJO phases, for ERA5 (top), one ACCESS-S2 ensemble member at lead week 1 (days 1-7) (middle left), GraphCast at lead week 1 (days 1-7) (middle right), one ACCESS-S2 ensemble member at lead week 3 (days 15-21) (bottom left)), and GraphCast at lead week 3 (days 15-21) (bottom right). The composites were constructed using rainfall data from periods where the MJO was active in that dataset (as opposed to using periods where the MJO was active in ERA5 for all composites). The number of cases with an active MJO contributing to each composite are indicated by the number in the upper right of each map.

rainfall i.e., it is not measuring forecast skill. Comparing the top and bottom panels reveals whether the models' ability to reproduce the rainfall patterns varies with forecast lead time. This analysis is particularly insightful for the ML model, as it helps assess its "physicality" - not only because the MJO index is derived from wind fields and not rainfall, but also because it evaluates the model's ability to reproduce large-scale, phase-dependent rainfall patterns rather than just grid-point level accuracy.

The rainfall composites show that for week 1, both GraphCast and ACCESS-S2 perform well in capturing both the spatial pattern and magnitude of rainfall anomalies associated with each MJO phase. By week 3, ACCESS-S2 continues to represent the spatial pattern with similar fidelity to week 1. In contrast, while GraphCast still captures the spatial pattern of the anomalies reasonably well, the magnitude of the response is too strong. One possible explanation could be related to the tendency of GraphCast (and other ML models trained against a mean squared error loss) to produce blurry outputs as lead-time increases (e.g. as shown in Husain et al., 2025), effectively removing smaller scale variability. This smoothing could suppress noise and inadvertently amplify coherent signals, leading to an over-emphasised MJO.

#### *3.1.3 The SAM*

The SAM is the leading mode of atmospheric variability in the Southern Hemisphere extratropics. It is characterised by meridional shifts in the strength of the zonal flow between about 55°–60°S and 35°–40°S. The positive phase of the SAM is associated with decreased geopotential height over Antarctica, increased geopotential height over the mid-latitudes and a poleward shift of the midlatitude westerly wind belt. Conversely, the negative phase of the SAM is associated with increased geopotential height over Antarctica, decreased geopotential height over the mid-latitudes and an equatorward shift of the midlatitude westerly wind belt (e.g., Thompson and Wallace 2000; Gillett et al. 2006). For Australia, the influence of SAM on rainfall and temperature is seasonally dependant. During austral winter, the positive phase of SAM is associated with decreased daily rainfall over southeast (i.e. Victoria and Tasmania regions) and southwest Australia. In contrast, during austral summer, the positive phase of SAM tends to enhance rainfall and decrease the incidence of extreme heat events along the southeastern coast of Australia (e.g. Hendon et al. 2007b). Current state-of-the-art physical prediction systems demonstrate skill in forecasting the daily SAM index out to about 14 days (e.g. Marshall et al. 2012; Wedd et al. 2022).

In order to assess the broad-scale depiction of the SAM in the models, and its representation with lead-time, the leading EOF of weekly-mean zonal-mean sea level pressure (SLP) between 25° and 75°S is calculated for ERA5 and each of the models over the period 1981-2018, following the methodology outlined in Marshall et al. (2012; Section 3.1). Figure 10 shows the first EOF for ERA5, as well as for lead weeks 1 and 3 from ACCESS-S2, FourCastNetV2 and GraphCast. The EOF1 pattern from all models closely resembles that of ERA5, accounting for about ~14% of the variability in weekly mean SLP. The SAM structure remains broadly consistent between week 1 and week 3 for all models, although GraphCast shows a slightly enhanced loading east of New Zealand at week 3.

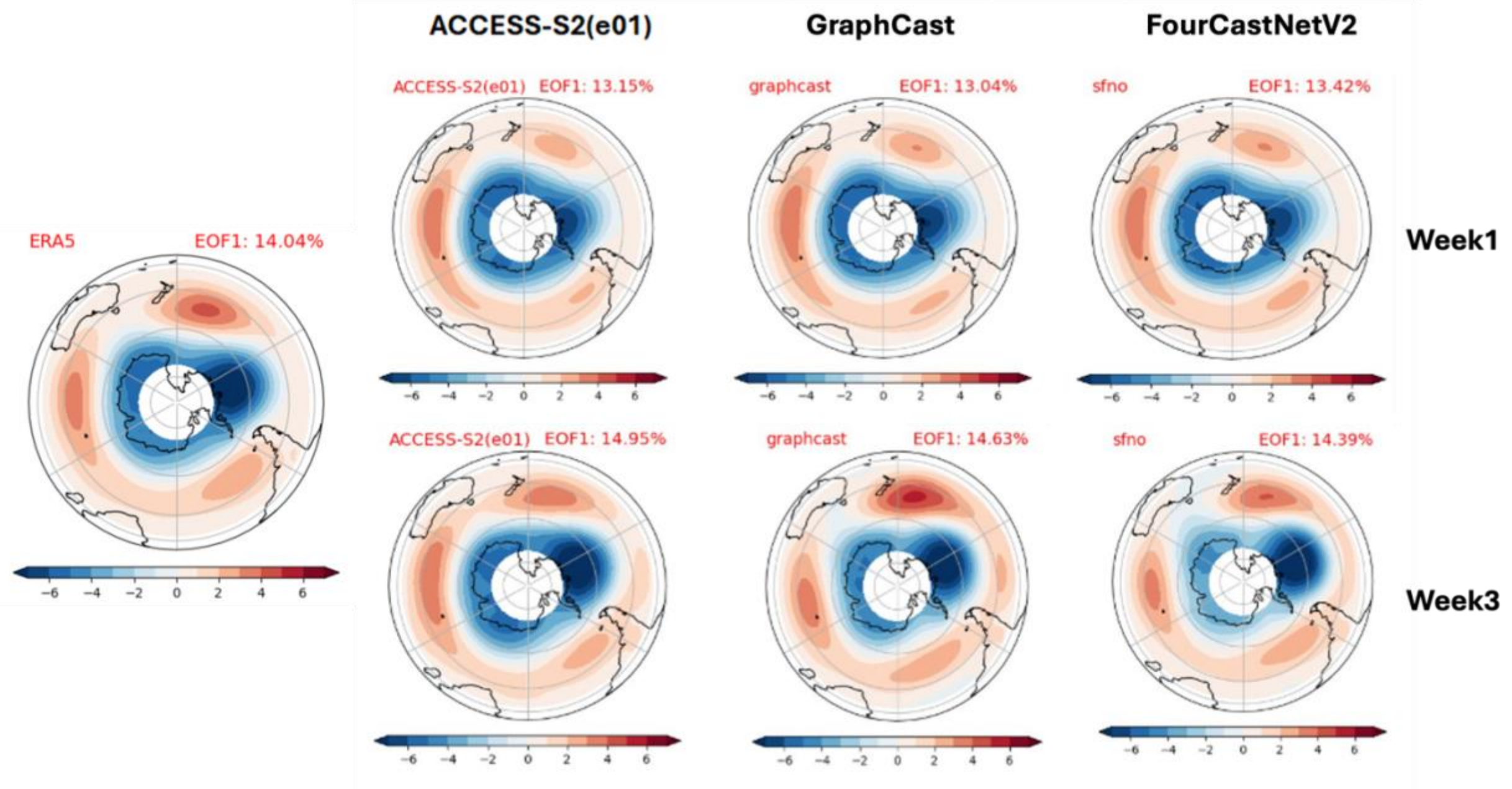

Figure 10: SAM patterns, here defined as the first EOF of the weekly mean (lead days 1-7 for week 1, lead days 15-21 for week 3) of the mean sea level pressure, for (a) ERA5, (b) ACCESS-S2 single ensemble member lead week 1, (c) GraphCast lead week 1, (d) FourCastNet V2 lead week 1, (e) ACCESS-S2 single ensemble member lead week 3, (f) GraphCast lead week 3, and (g) FourCastNet V2 lead week 3. Weekly means are used to compute the EOF instead of the more common monthly mean because the forecast models have lead times shorter than a full month, making monthly means impossible. The EOF patterns from the weekly means are very similar to those obtained by using the monthly mean.

To evaluate the skill in forecasting the daily SAM index, daily mean predicted SLP anomalies are projected onto the observed (ERA5) EOF, rather than using each model's own EOF. Specifically, the anomalies are projected onto the first EOF of ERA5 monthly mean SLP between 25° and 75°S, and then normalised by the standard deviation of the observed (ERA5) monthly SAM index (see Marshall et al. 2012 for details). The correlation and RMSE for forecasts of the daily SAM index are shown in Figure 11. The two ML models exhibit comparable skill in predicting the SAM index. At shorter lead-times (up to ~10 days), they perform slightly better than the ACCESS-S2 ensemble mean and outperform individual ACCESS-S2 ensemble members. At longer lead-times, the ensemble mean of ACCESS-S2 demonstrates superior skill; however, both GraphCast and FourCastNetV2 are similar to, if not a bit better than, individual ACCESS-S2 ensemble members. This highlights the possibility that an ensemble ML model may surpass the skill of the ACCESS-S2 ensemble mean at longer lead-times.

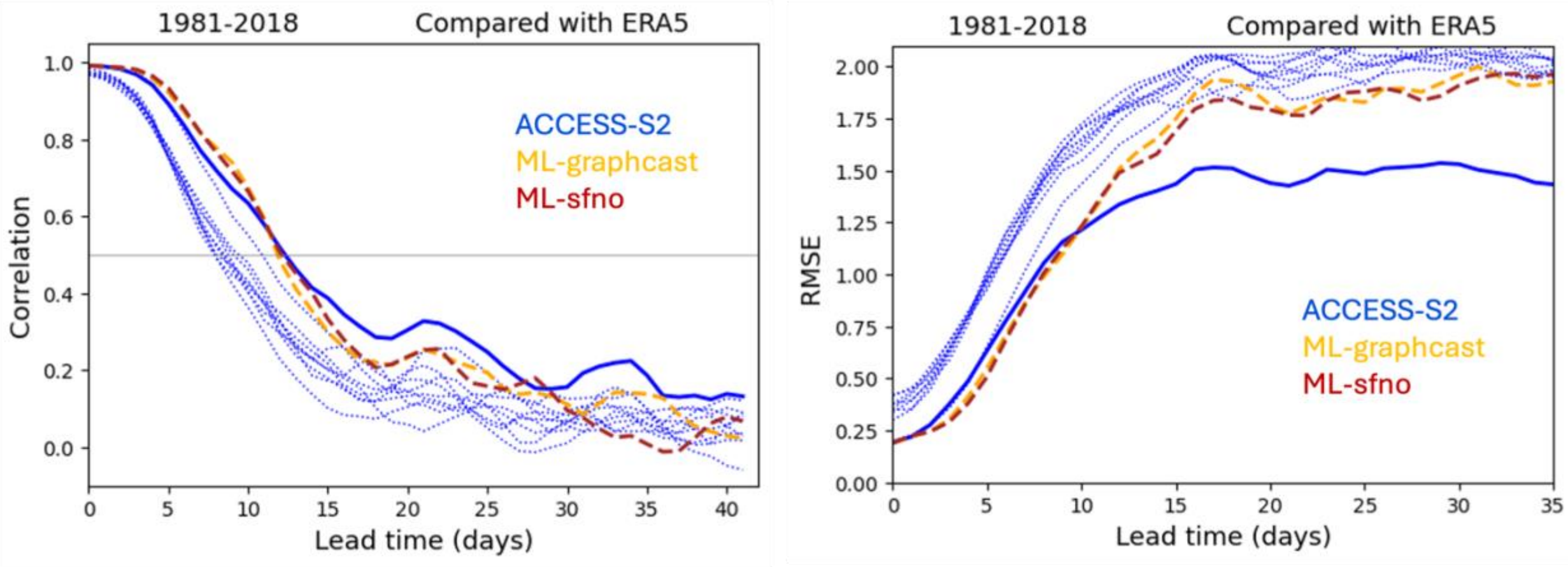

Figure 11: Daily SAM index anomaly correlation (left) and RMSE (right) skill for all months in all years from 1981 to 2018 for ACCESS-S2 ensemble members (dotted blue), the ACCESS-S2 ensemble mean (solid blue), GraphCast (dashed yellow) and FourCastNetV2 (dashed red). The daily SAM index is obtained by projecting the sea level pressure anomaly of the southern hemisphere (75S-25S) to the observed EOF1 pattern, and normalizing by the observed standard deviation of the monthly SAM index.

## 3.2 The 2.5-year hindcast period

This section presents an additional analysis with the 2.5-year hindcast that is independent of the ML models training periods. The physical model used here as a benchmark is the more recent global coupled model configuration, GC5, rather than the GC2 model used in ACCESS-S2. The focus is to highlight any key differences from the results in the previous section, while also identifying areas of consistency to build confidence in findings across both hindcast periods. Where possible, the assessment methods from the previous section were repeated. However, a major limitation of this shorter period is the inability to compute anomalies with respect to the model's climatology. Using an ML model climatology from a longer period overlapping with the ML model training windows would compromise data independence, and the 2.5-year period alone is too short to establish a reliable climatology. As a compromise, all anomalies are calculated with respect to the ERA5 climatology (1981-2018). Since the anomalies are not with respect to the model's lead-time dependent climatology, model biases (including lead-time dependent biases) are not accounted for, and it is therefore likely that the results shown here are an underestimate of what the model performance would be without this limitation. In addition, given the very short analysis period, the sample size of events (e.g. MJOs of a given phases) is very small. This means that there is considerable noise in the results compared to the 38-year analysis.

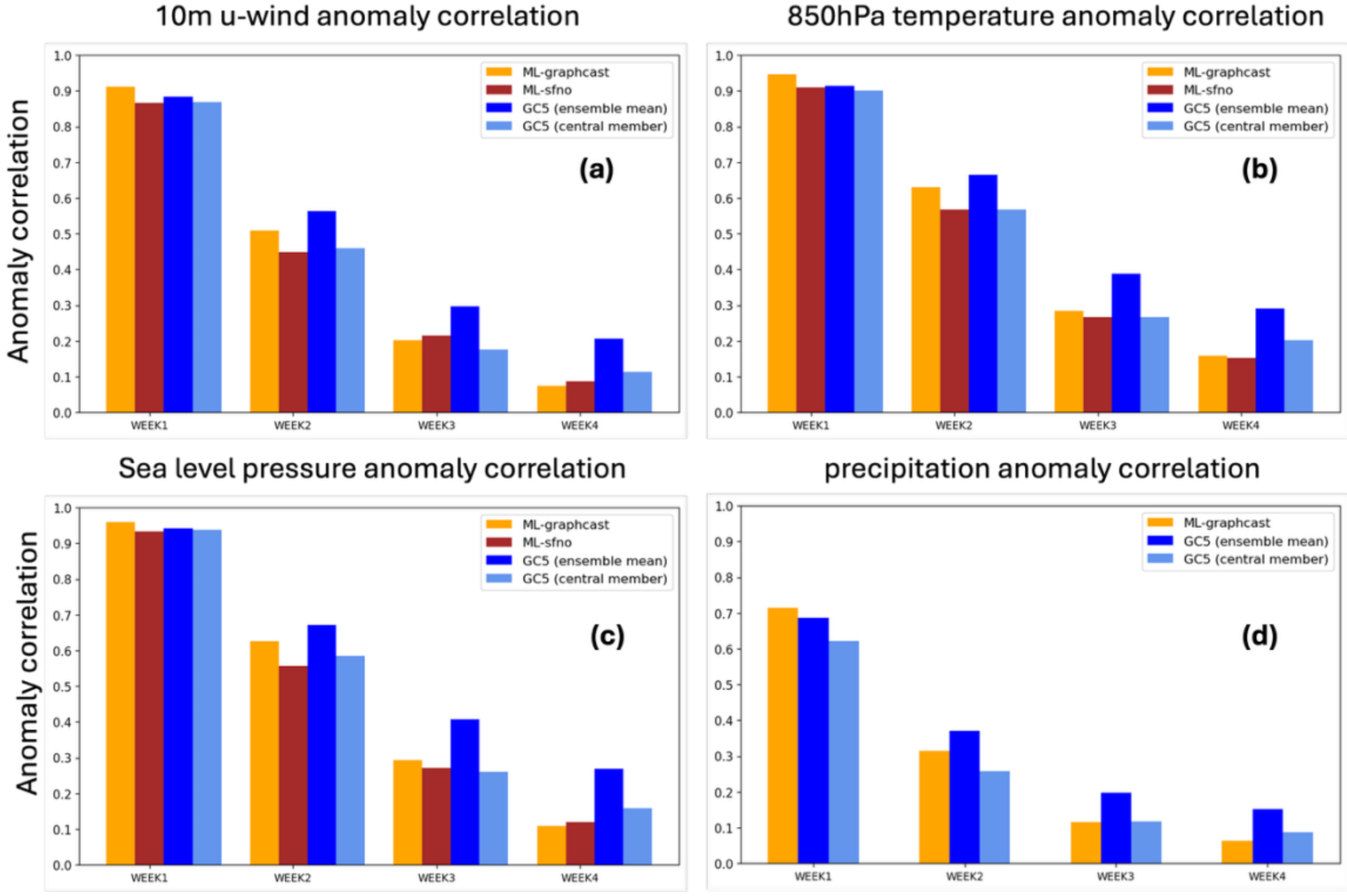

Figure 12: Globally averaged correlation skills of (a) 10m u-wind speed, (b) 850 hPa temperature, (c) mean sea level pressure, and (d) rainfall for lead weeks 1 to 4 for the 2.5-year hindcast. Yellow, red, dark blue and light blue bars show the correlations for GraphCast, FourCastNetV2, the 9-member GC5 ensemble mean and a single member of GC5 respectively. All anomalies are defined relative to the ERA5 climatology for the period 1981-2018.

### *3.2.1 Weekly skill*

Globally-averaged weekly correlation skills of 10m u-wind speed, 850 hPa temperature, mean sea level pressure, and rainfall for weeks 1 to 4 for the 2.5-year hindcast period are presented in Figure 12. The broad patterns in correlation skill with lead week are consistent with those in the 38-year hindcast period, although here GC5 tends to show slightly higher correlation skills compared to the ML models (particularly for week 2) than ACCESS-S2 did in the 38-year hindcast period. This is not surprising since ACCESS-S2 is built on the GC2 model, several model versions behind GC5. In addition, GC5 was run at a higher spatial resolution (~40km) compared to ACCESS-S2 (~60km). Nonetheless, the main source of additional skill in GC5 over the ML models is in the ensemble mean, reinforcing the conclusion that there is value in exploring the impact of ensemble methods for ML models alongside ensemble physical model systems.

Mirroring those in the previous section, maps of the correlation skill for weekly mean screen-level temperature and precipitation anomalies for weeks 1 and 3 are shown in Figures 13 and 14. As with ACCESS-S2 in the previous section, here the GC5 maps are for a single ensemble member, not the ensemble mean. It is clear that due to the short hindcast period, the maps are noisier than those from the 38-year period (Figures 2 and 3). In addition, the results will be highly sensitive to the drivers active during the 2.5-year hindcast period and are probably significantly influenced by individual weather events. Therefore, caution should be exercised in over-interpreting the results in comparison with the 38-year period.

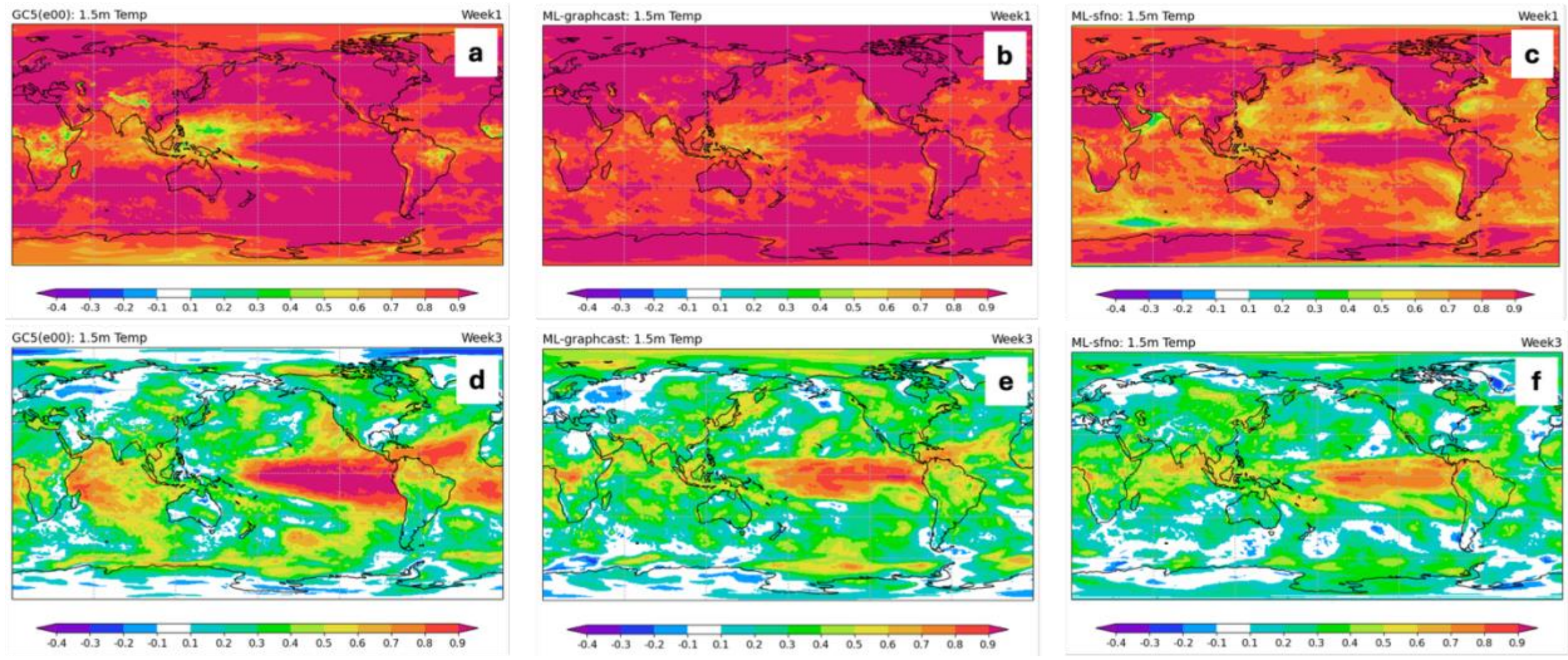

Figure 13: The upper-row figures illustrate the anomaly correlation skills of weekly temperature at 1.5m at lead time week 1 across the 2.5-year hindcast period for: a single GC5 ensemble member (a), GraphCast (b), and FourCastNetV2 (c). The bottom-row figures (d–f) represent the same variables as the upper row but at lead time week 3. All anomalies are defined relative to the ERA5 climatology for the period 1981-2018.

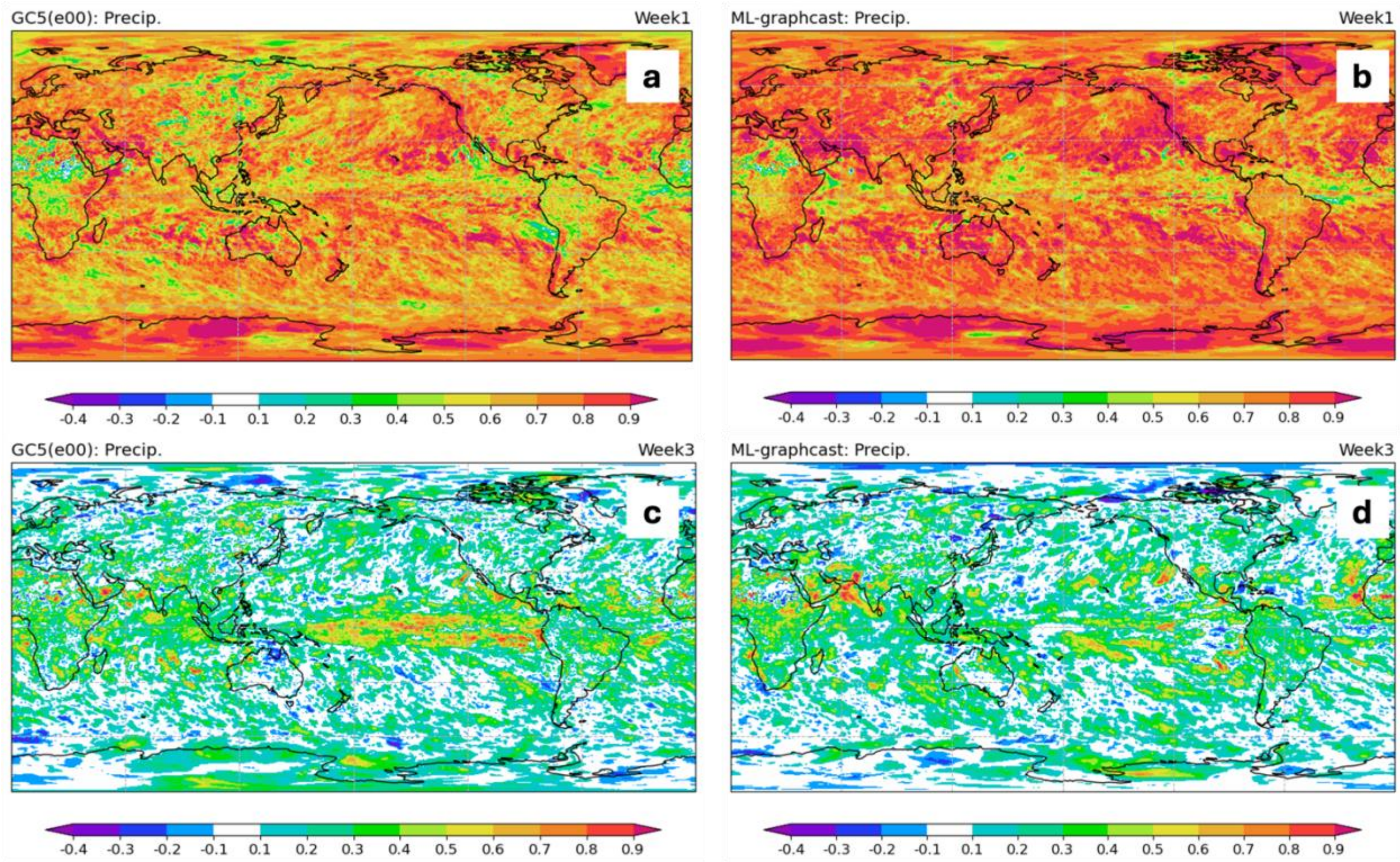

Figure 14: As in Figure 13 except for precipitation.

For screen-level temperature at week 1, as was seen for the 38-year period results, the ML models both show comparable or higher skill over the land compared to the physical model (Figure 13). Over the oceans, for GraphCast and GC5, the skill is relatively higher for the 2.5-year period (although GC5 does perform worse over equatorial Africa and the eastern Maritime Continent including the western equatorial Pacific), whereas for FourCastNetV2 the extratropical oceans are somewhat less skilful in the 2.5-year period compared to the 38-year period, and FourCastNetV2 has the lowest overall skill over the oceans of all three models.

At week 3, as was seen for the 38-year period, the physical model has superior skill over the oceans, which is not surprising given it is the only model to feature an ocean component. While the broad patterns of high and low correlations are consistent between the three models, the ML models feature more areas of lower correlations, such as in the Indian ocean and mid-latitudes of the Atlantic Ocean.

As expected, the maps of skill for rainfall (Figure 14) are noisier than those for temperature. In the 2.5-year period, the skill that emerges at week 3 over the equatorial Pacific for GraphCast is potentially less prominent with respect to that from the physical model than it was in the 38-year period (Figure 3).

While there is more noise in the correlation maps here compared to those from the 38-year hindcast, in general, the spatial distribution of high and low correlations are similar, as is the degree of change in the correlation skills from week 1 to week 3. However, it is also clear that the small sample size makes it challenging to draw any robust conclusions.

#### *3.2.2 The MJO*

The RMM indices for the 2.5-year hindcast are computed in the same way as for the 38-year hindcast, except that anomalies are calculated using the ERA5 climatology from 1981 to 2018, instead of the model's own climatology as was done for the 38-year hindcast period. As such, systematic model biases are not removed from the anomalies by this procedure.

Two cases (chosen arbitrarily) from the 2.5-year hindcast period are presented in Figure 15, with forecasts initialised on the 1st of March 2023, and the 1st of March 2024 for GC5, GraphCast and FourCastNetV2. In each case the forecasts had a lead-time of 28 days.

Considering the MJO phase diagram of the 2023 case, the MJO event began in the western Pacific, intensified as it propagated across the Pacific and Atlantic regions, and began to weaken before entering the Indian Ocean region. All models captured the intensification of the MJO in the Pacific and its eastward propagation in the first two weeks, as shown in both phase diagram and zonal wind anomalies.

In the March 2024 case, where the MJO is already active at forecast initialization and moves from the Maritime Continent to the Western Pacific, the two ML models appear to do a better job in capturing the magnitude of MJO than GC5, although GraphCast significantly underestimates the propagation speed compared to ERA5 and the other models. These two cases highlight that the relative performance of the ML models and GC5 in predicting the MJO varies depending on the specific cases, but that in general the ML models are capable of producing a reasonable MJO.

The RMSE and correlation skill of the models for predicting the MJO over the 2.5-year hindcast are shown in Figure 16. The GC5 ensemble mean has correlation skills more similar to that of the individual ensemble members than was seen in the 38-year hindcast (Figure 5). This may be due to the different ensemble generation methods used for GC5 and ACCESS-S2 in this study, resulting in less spread in the ensemble members in the tropics using the GC5 system and therefore less added benefit of an ensemble mean. ACCESS-S2's perturbation of the atmosphere initial conditions is focussed on optimising for the sub-seasonal timescale (Wedd et al., 2022), whereas the experiments with GC5 use the MOGREPS weather forecast initial perturbations (Bowler et al., 2008). FourCastNetV2 has skill similar to GC5 over the first week of the forecast and is competitive with individual ensemble members thereafter. The amplitude prediction skill of the models in the 2.5-year hindcast period (not shown) has similar properties to the 38 year hindcast period (Figure 6), with GraphCast overpredicting the magnitude of the MJO signal beyond two weeks, and ACCESS-S2 and FourCastNetV2 predicting it more skilfully with a small underestimation relative to ERA5.

The most obvious difference between the results of the 38-year period and the 2.5-year period is the much poorer performance of GraphCast in the latter, relative to the other models. This is also consistent with the low and negative correlations seen for GraphCast in the vicinity of the Maritime Continent in Figure 13. GraphCast suffers from relatively large wind biases in the Maritime Continent region relative to both ACCESS-S2 and FourCastNetV2 in the 38-year period (Figure 7) and compared to GC5 and FourCastNetV2 in the 2.5-year period (Figure 17). In the 38-year period, these model-dependent biases are removed in the calculation of the anomalies and GraphCast MJO forecast skill is competitive to that of FourCastNetV2 and ACCESS-S2. However, for the 2.5-year period, when the ERA5 climatology is used to create the anomalies (and the lead-time dependant model bias is not removed), this has a larger effect on GraphCast than the other models. Additionally, despite the limited sample size, we can observe the slow MJO propagation in GraphCast (Figure 18) for phases 1 and 3, consistent with the 38-year climatological results, and conversely, propagation speeds close to those of events in ERA5 for phase 2, which is not consistent with the 38-year climatological results. Clearly these observations should be treated with caution given the very limited number of MJO cases available in this period (e.g., there were no GC5 MJO cases in the sample that started in phase 1 and remained active for at least 5 days). We have not repeated the MJO rainfall composites for the 2.5-year hindcast period because there were insufficient MJO events.

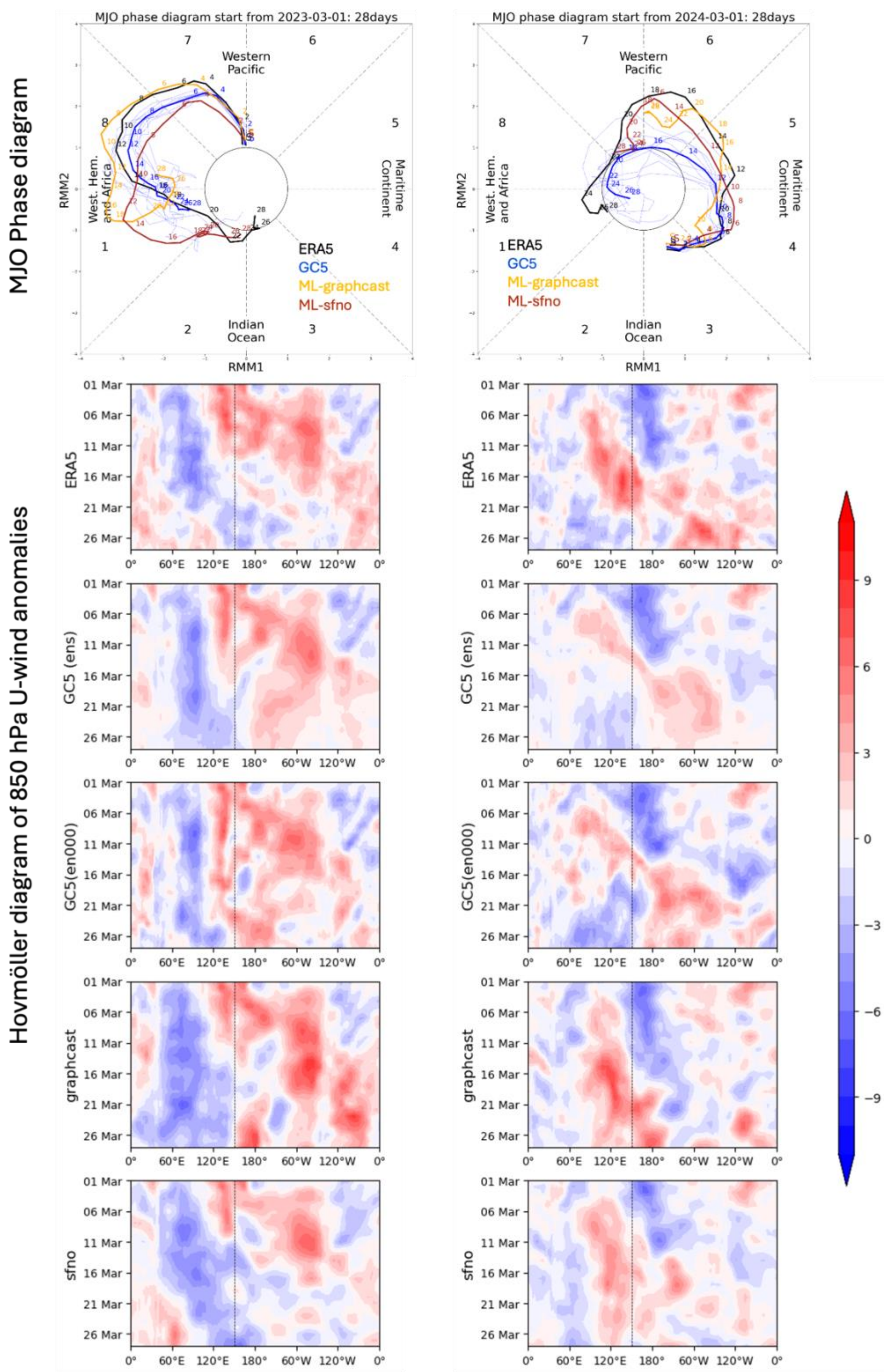

Figure 15: The phase diagram (top row) for two MJO events, beginning from the 1st of March 2023 (left column) and the 1st of March 2024 (right column), and the corresponding U850 anomalies evolution for each event in (from top to bottom) ERA5, the GC5 ensemble mean, one GC5 ensemble member, GraphCast, and FourCastNetV2.

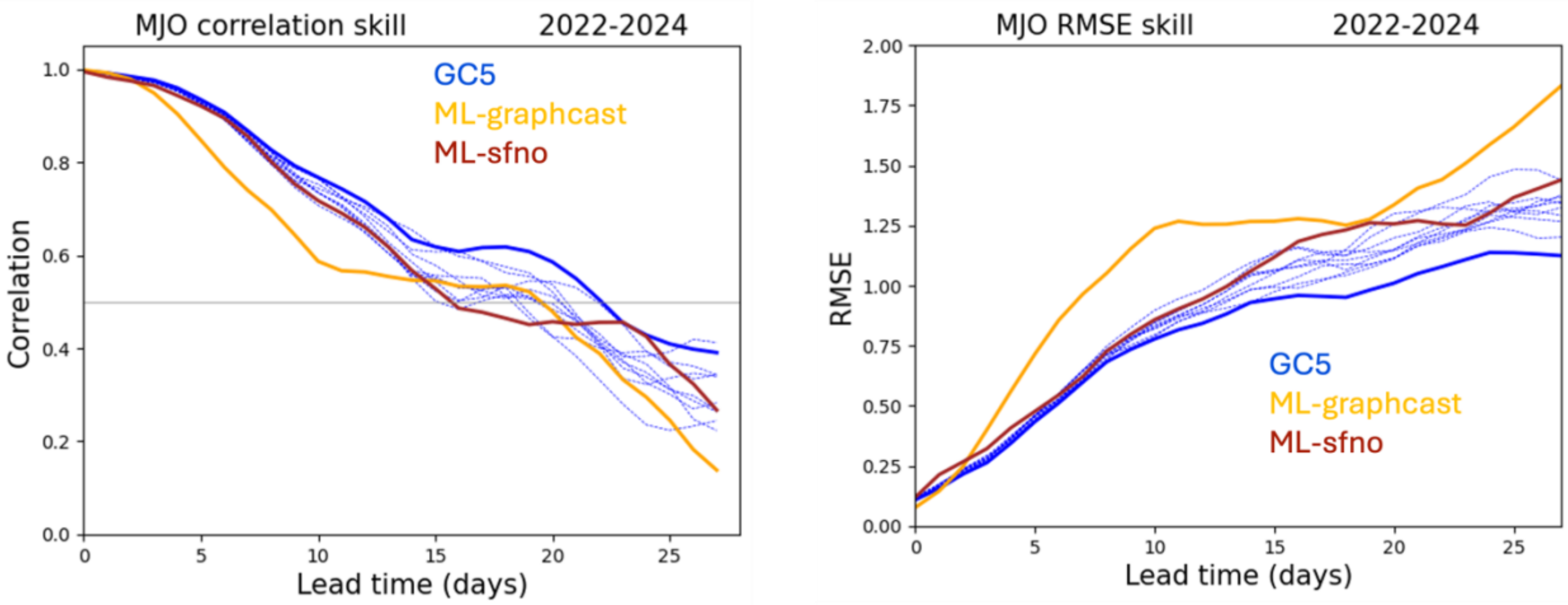

Figure 16: MJO anomaly correlation (left) and RMSE (right) skill for all months in all years from Jan 2022 to Jun 2024 for GC5 ensemble members (dotted blue), the GC5 ensemble mean (solid blue), GraphCast (yellow) and FourCastNetV2 (red). Anomalies are calculated with respect to the ERA5 climatology (1981-2018) rather than the model's own climatology.

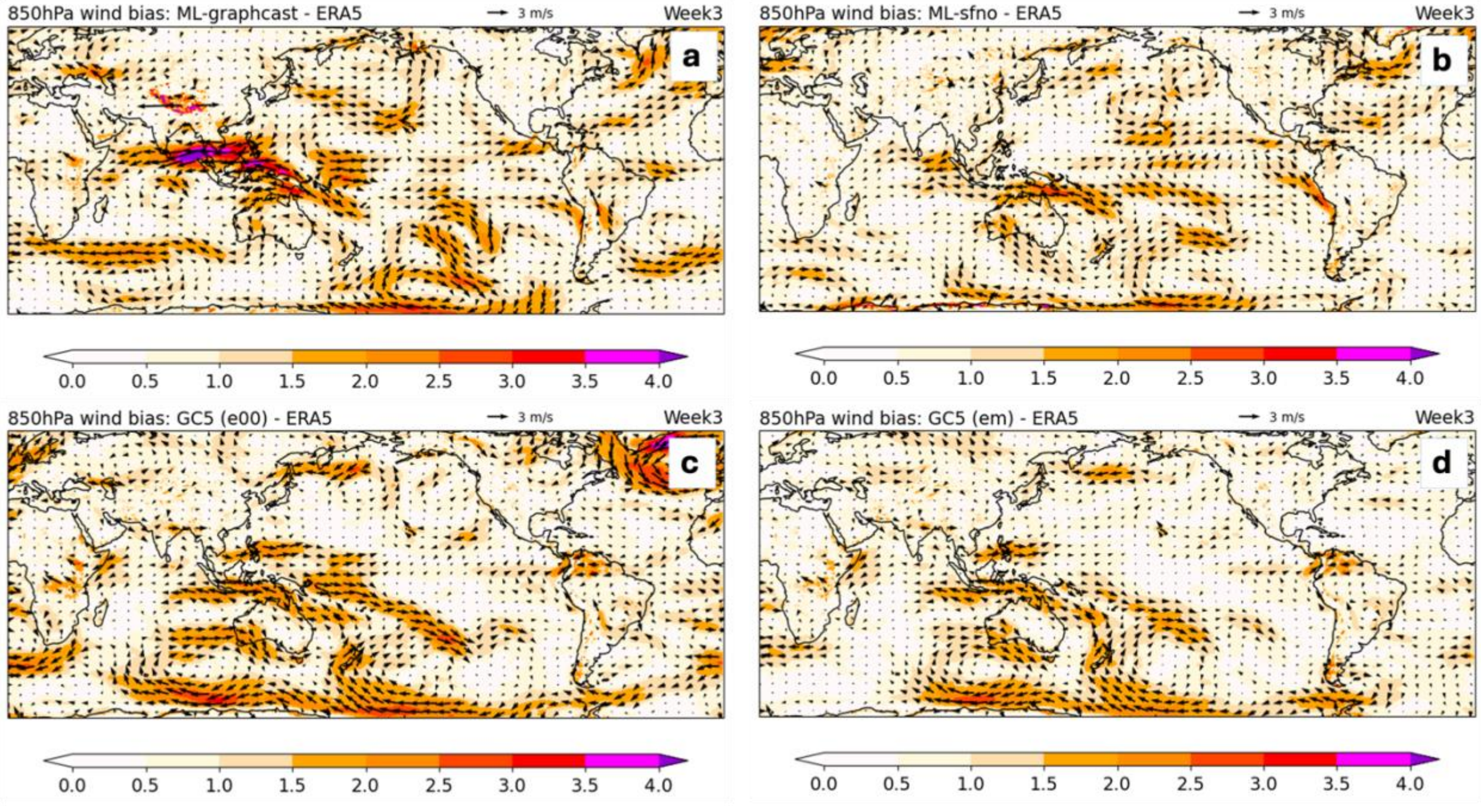

Figure 17: Weekly-average 850 hPa annual mean wind biases relative to ERA5 across the 2.5-year hindcast period for (a) GraphCast, (b) FourCastNetV2, (c) one GC5 ensemble member, and (d) the GC5 ensemble mean, at lead week 3. The biases are calculated by computing weekly means along lead time, followed by annual means across model runs, and then subtracting from the equivalent from ERA5, before averaging the difference across the 2.5-year hindcast period.

### *3.2.3 The SAM*

Correlation and RMSE skills for the SAM index are presented in Figure 19. The conclusions for the performance of the SAM from the 2.5-year period are broadly similar to that from the 38-year period - that the ML models have performance similar to the ensemble mean of the physical model at shorter lead-times and are generally comparable with the ensemble members thereafter. Interestingly, the RMSE for GC5 is unexpectedly high for the first ~3 days compared to the ML models. This is likely due to the larger sea level pressure biases present in GC5 at southern hemisphere high latitudes at early lead-times (not shown), which could impact the SAM index in the 2.5-year hindcast period since the sea level pressure anomalies were calculated relative to an ERA5 climatology rather than the model's own climatology.

As with the MJO rainfall composites, we have not repeated the SAM EOF calculations for the 2.5-year hindcast period because there were insufficient samples.

# 4 Discussion

The investigation of skill in both the 38-year and 2.5-year hindcasts showed that GraphCast and FourCastNetV2 are broadly competitive with the ensemble means of both ACCESS-S2 and GC5, respectively, at short lead-times (<2 weeks), and are competitive with the physical model ensemble members and inferior to the physical model ensemble means at longer lead times (>2 weeks). This is clear from both the globally averaged results and the skill for forecasts of the SAM and MJO. It shows that the models are relatively stable out to longer leads, and their performance is impressive given that they were not developed for sub-seasonal prediction.

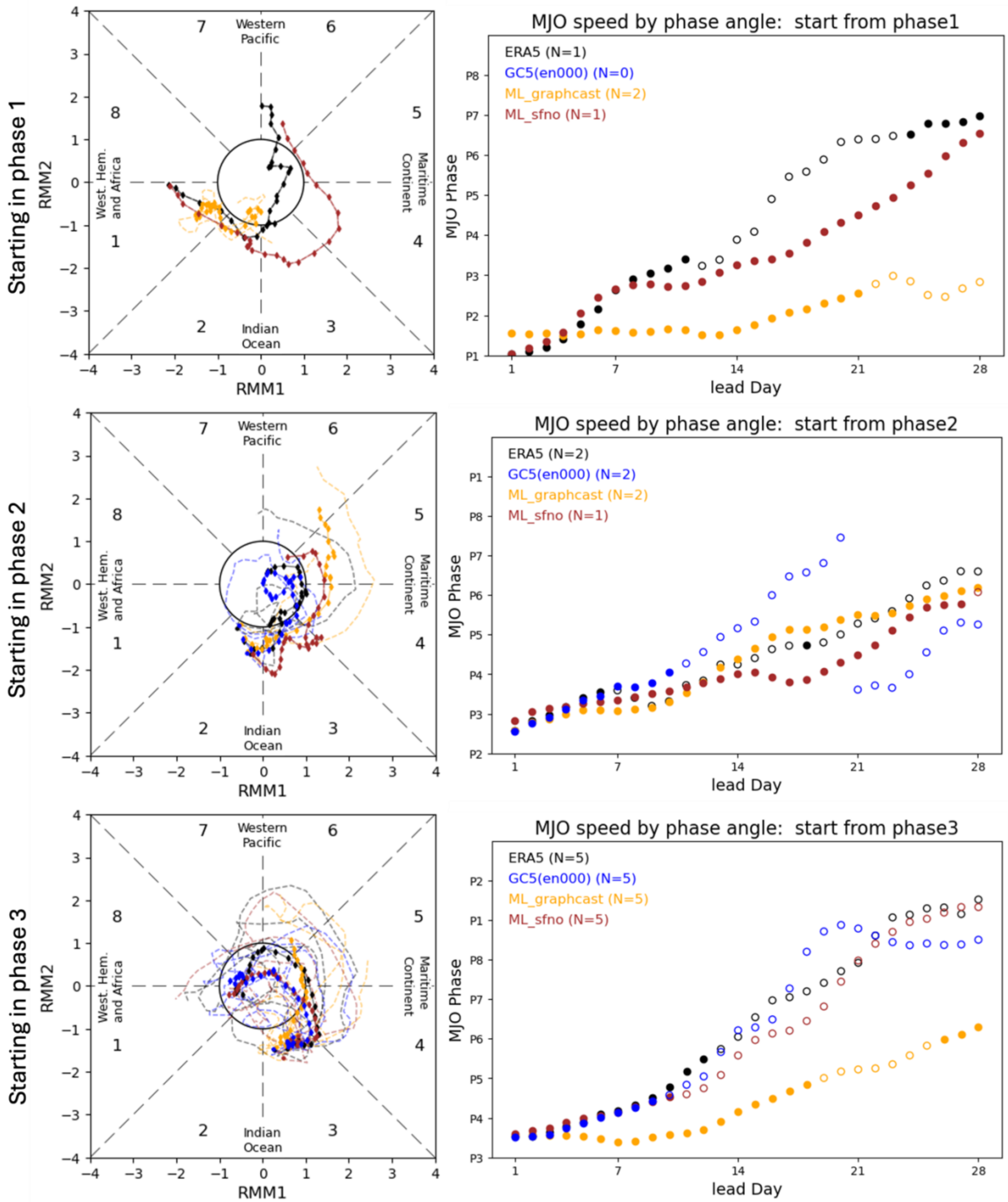

Figure 18: Left Column: Phase diagrams for all MJO events in the 2.5-year hindcast period active for more than 5 days starting in phase 1 (top), phase 2 (middle) and phase 3 (bottom) for GC5 (blue), GraphCast (yellow), FourCastNetV2 (red), and ERA5 (black). A threshold of being active for 5 days was used instead of the 14 days used for the 38-year hindcast period due to the limited number of MJO cases in the 2.5-year hindcast period. The dashed lines show individual MJO events, and the solid lines show the composite of all events. Right column:

Composite phase speed plots for all MJO events in the 2.5-year hindcast period active for more than 5 days starting in phase 1 (top), phase 2 (middle) and phase 3 (bottom) for GC5 (blue), GraphCast (yellow), FourCastNetV2 (red), and ERA5 (black). On this plot, a faster propagating MJO event has a steeper slope, and a slower event has a shallower slope. Open circles indicate a composite amplitude that is weak (less than 1) while solid circles indicate a composite amplitude greater than 1. The number of cases contributing to the figure for each dataset is indicated in the legend of the right column plots. No MJO cases in GC5 met the criteria in phase 1 (top row).

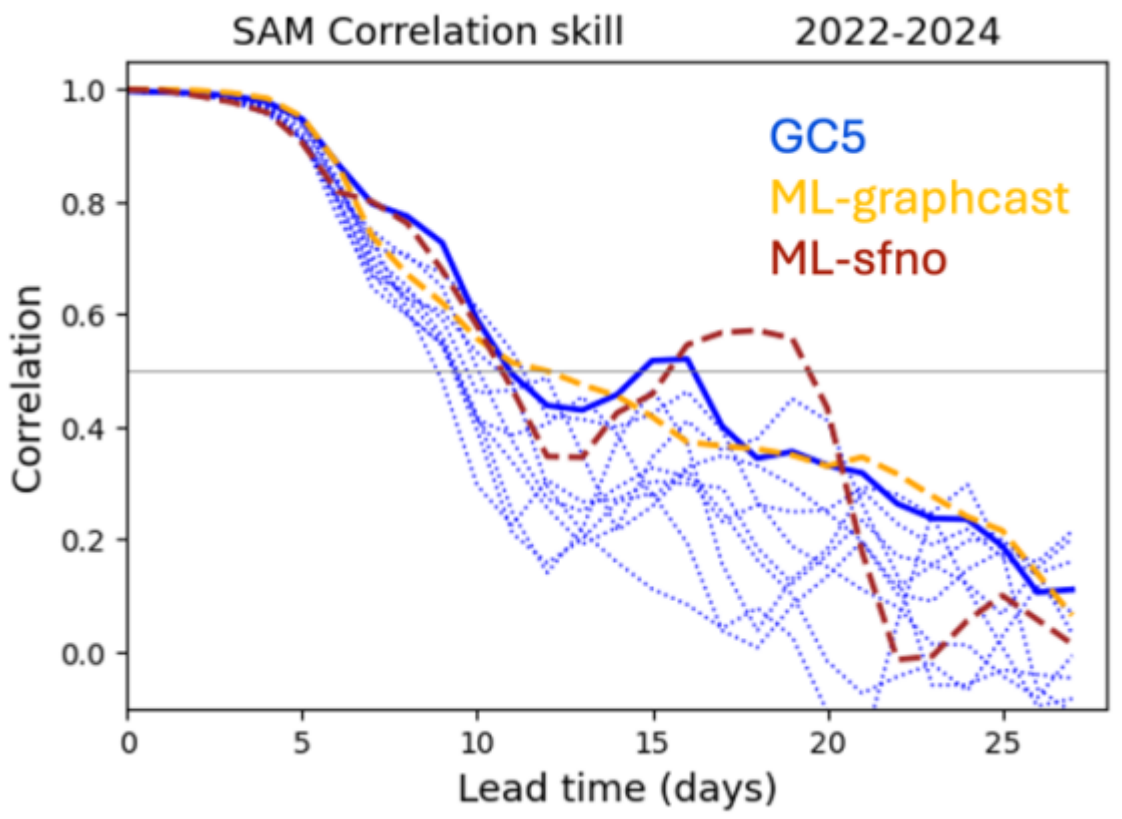

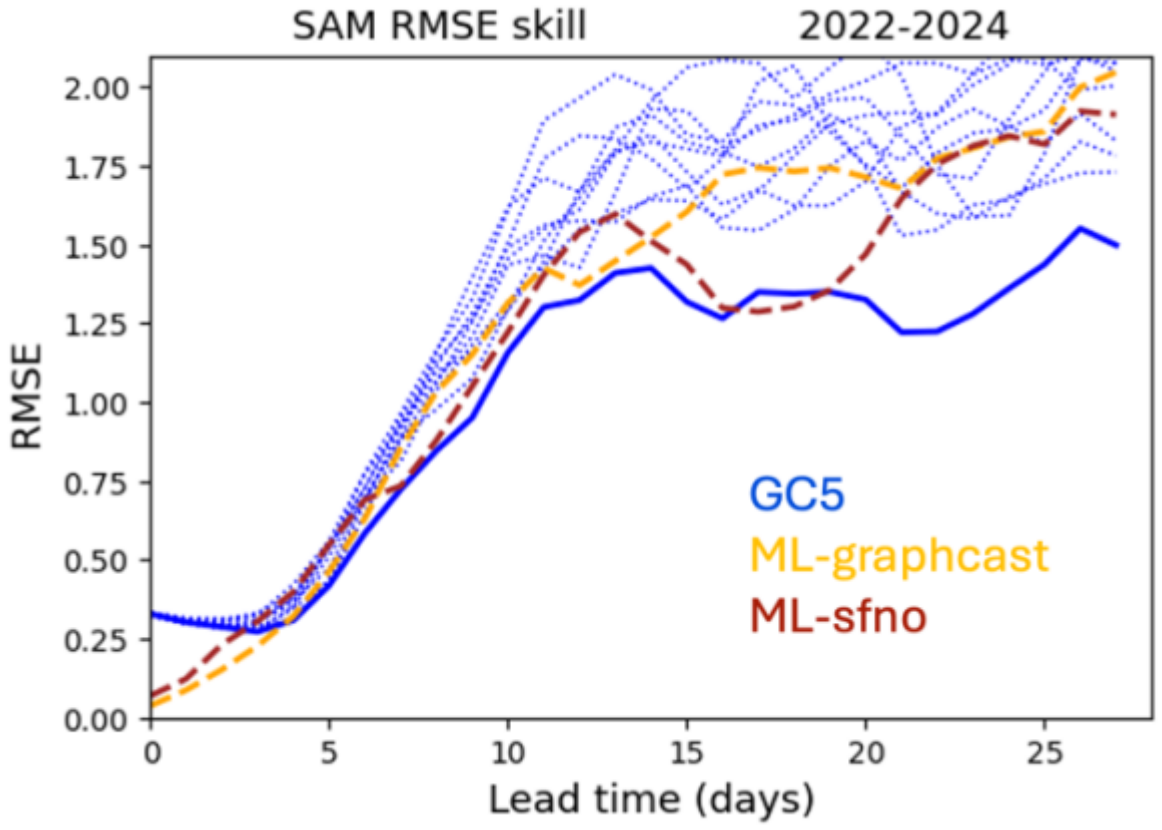

Figure 19: Daily SAM index correlation (left) and RMSE (right) skill for all months in all years from Jan 2022 to Jun 2024 for GC5 ensemble members (dotted blue), the GC5 ensemble mean (solid blue), GraphCast (yellow) and FourCastNetV2 (red). The daily SAM index is obtained by projecting the sea level pressure anomaly of the southern hemisphere (75S-25S) relative to the ERA5 climatology to the observed (ERA5) EOF1 pattern and normalizing by the observed (ERA5) standard deviation of the monthly SAM index. The climatology, EOF1 pattern, and standard deviation of the monthly SAM index of ERA5 were all calculated over the period 1981-2018.

However, the spatial maps indicate that there are key deficiencies in the ML models for sub-seasonal prediction. At week 3 of the forecasts, whilst the ML models indicate similar regions of high skill compared to the physical models, their skill over the tropical and midlatitude oceans is significantly lower than that from the physical models (Figures 2, 3, 13, and 14). This probably underscores the lack of ocean variables as predictors for the ML models, which are important for sub-seasonal prediction and become increasingly important as lead-time increases. The lack of ocean variables and representation of coupled ocean-atmosphere processes may also be playing a role in the relatively lower skill in the ML models, and GraphCast in particular, in the region of the Maritime Continent.

Despite not having variables from the ocean sub-surface, and even though coupling with the ocean is thought to be important for the MJO, the results from the 38-year and 2.5-year hindcasts indicate that the ML models have good skill for forecasts of the MJO. GraphCast does, however, overestimate the amplitude of events beyond the first week in the 38-year hindcast (Figure 6) and the 2.5 year hindcast (not shown), possibly due to large U850 wind biases over the Maritime Continent. The case studies (including those in the 2.5 year hindcast outside the ML model training period) show that the ML models can simulate reasonable looking MJOs, which are not distinguishable as being from ML models without prior knowledge (Figures 4 and 15). Rainfall composites from the 38-year hindcast (Figure 9) also show that GraphCast effectively captures the relationship between MJO phase and tropical rainfall, indicating its capability to represent broad-scale atmospheric relationships. However, the composite values are overly strong at longer leads (e.g., week 3), potentially reflecting a tendency of the model to smooth fields with increasing lead-time (as shown in Husain et al., 2025). More recent models like AIFS-CRPS (Lang et al., 2024), GenCast (Price et al., 2025), FuXi-ENS (Zhong et al., 2025), and FourCastNetV3 (Bonev et al., 2025) exhibit less smoothing, due to a variety of factors including: use of diffusion and other generative architectures; loss functions which reward accurate prediction of the whole forecast distribution like CRPS (Continuous Ranked Probability Score) instead of metrics focussed on the middle of the distribution like RMSE; and injection of noise into the model at each timestep. It would be interesting to examine if these models produce tropical rainfall patterns with more realistic amplitudes at extended lead-times.

The results of this study also raise the question of the extent to which the skill of the ML models examined could be improved by an ensemble forecast. While GraphCast and FourCastNetV2 were not designed to produce an ensemble forecast, AIFS-CRPS was (as were GenCast, FuXi-ENS, and FourCastNetV3), and results show that when forecasting to sub-seasonal lead-times (including for the MJO) it outperforms the ECMWF physical model when dealing with full fields (i.e. when not removing the bias) and is comparable when forecasting anomalies (Lang et al., 2024). Since they do not natively produce ensemble outputs, constructing an ensemble with GraphCast and FourCastNetV2 would require traditional approaches like initial condition perturbation or time-lagged ensembles. However, while such strategies have been explored for ML models on weather timescales (e.g., Bi et al., 2023), very few end up with well-calibrated ensembles. It remains unclear to what extent it is beneficial to ensure ensemble perturbations and spread are flow-dependent to better reflect the spatial and temporal characteristics of uncertainty, rather than just satisfying bulk statistical properties.

One of the key advantages of ML models is their significantly lower computational cost and faster inference time. This makes them particularly attractive for sub-seasonal and seasonal prediction, where generating hindcasts can be completed orders of magnitude more quickly and cheaply than with traditional methods. Additionally, unlike physical models, ML models (in theory) do not need to accurately simulate every intermediate physical process linking a climate driver - such as the El Niño–Southern Oscillation - to its teleconnected impacts. This opens the possibility that ML models could still capture downstream effects even when the intermediate processes are poorly represented in the model (or even the training data). In other words, ML models can potentially learn both the teleconnections and the underlying physical processes independently, rather than relying on teleconnections to emerge from the modelled physics. While the analysis presented here does not directly address this hypothesis, it provides a foundation for future studies that aim to explore these questions in greater depth.

There was reasonable agreement between the assessments undertaken in the 38-year and 2.5-year hindcast periods in the high-level conclusions, providing greater confidence in the results, and demonstrating the value of our approach to mitigating the limited independent evaluation period. However, it is also clear that the 2.5-year period is too short to obtain robust conclusions about sub-seasonal prediction. There are significant challenges around having a large enough sample to create climatologies (to enable model bias correction) and having enough samples of events, such as different phases of the MJO. As outlined in the Introduction, a key challenge in evaluating ML models for sub-seasonal, seasonal, and longer timescales is the limited amount of independent data available for evaluation after training. While this is generally sufficient for weather timescales, it becomes inadequate at extended timescales. Our compromise approach - comparing results across two hindcast periods - relies on the assumption that agreement between them reflects robustness in the results. This is of course not without limitations, since any apparent agreement may be due to a coincidental alignment between the 2.5-year sample and the 38-year sample. Furthermore, the inability to remove a model-dependent climatology in computing the anomalies for the 2.5 year period, meaning biases are not accounted for, makes direct comparison to the 38-year hindcast analysis more challenging. This method is therefore not intended as an optimal solution, but rather a temporary workaround for cases where insufficient evaluation data has been retained. A more robust alternative would be to reduce the training period to preserve a longer evaluation window, although this becomes increasingly impractical as the required verification period grows with forecast lead time. This underscores a fundamental challenge facing the community as ML models are applied to extended-range and multi-year prediction.

# 5 Conclusions

To gain an initial insight into the performance of ML models for sub-seasonal prediction, two medium range ML weather models, GraphCast and FourCastNetV2, have been investigated. Despite not being designed for sub-seasonal prediction and lacking ensembles, they have been found to demonstrate surprisingly good performance at longer lead times and in capturing key drivers of climate variability, specifically the MJO and the SAM.

A dual-hindcast approach has been explored, involving the use of a long hindcast period to obtain robust results, though this overlaps the ML model training data, alongside a shorter, but independent hindcast set. Assessing the degree of agreement (or disagreement) between the two provides a measure of confidence in the trustworthiness of the overall evaluation. Although this strategy is not as effective as a single long and independent dataset, it offers a practical solution considering the limited independent data available for evaluation for many ML weather models. Looking ahead, it is recommended that, as much as possible, developers of ML models targeting sub-seasonal and longer timescales retain sufficient data outside the training window to enable robust evaluation.

The assessment presented here lays the groundwork for future research focussed on ML models specifically designed for sub-seasonal prediction – those that include an ensemble approach and consider variables linked to low-frequency predictability, such as those from the ocean, stratosphere and land surface. Further evaluation should also investigate the capability of ML models to predict extreme and high-impact sub-seasonal events.

ML models are rapidly transforming weather and climate prediction, offering exciting opportunities and new challenges. More work establishing baseline evaluations of these models against physical models will be important going forwards.

# Acknowledgements

We thank many staff across the Bureau for their efforts, advice and feedback contributing to this work. We thank Matthew Wheeler, Hanh Nguyen, and Rajashree Naha for valuable review of this manuscript prior to submission. We also thank ECMWF for providing invaluable datasets such as ERA5 to the research community. This work was partly funded by Meat & Livestock Australia, the Queensland Government through the Drought and Climate Adaptation Program, and the University of Southern Queensland through the Northern Australia Climate Program (NACP).

# Data Availability

ACCESS-S2 data are published on the National Computational Infrastructure (NCI) and are available for research purposes[1]. GraphCast, FourCastNetV2 and GC5 data used in this study are available on request.

[1] https://geonetwork.nci.org.au/geonetwork/srv/eng/catalog.search#/metadata/f3311_4920_0252_8073

# References

Alet, F., Price, I., El-Kadi, A., Masters, D., Markou, S., Andersson, T. R., ... & Battaglia, P. (2025). Skillful joint probabilistic weather forecasting from marginals. arXiv preprint arXiv:2506.10772.

Antonio, B., Strommen, K., & Christensen, H. M. (2025). Seasonal forecasting using the GenCast probabilistic machine learning model. arXiv preprint arXiv:2509.06457.

Ben Bouallègue, Z., Clare, M. C., Magnusson, L., Gascon, E., Maier-Gerber, M., Janoušek, M., ... & Pappenberger, F. (2024). The Rise of Data-Driven Weather Forecasting: A First Statistical Assessment of Machine Learning–Based Weather Forecasts in an Operational-Like Context. Bulletin of the American Meteorological Society, 105(6), E864-E883.

Bi, K., Xie, L., Zhang, H., Chen, X., Gu, X., & Tian, Q. (2023). Accurate medium-range global weather forecasting with 3D neural networks. Nature, 619(7970), 533-538.

Bonavita, M. (2023). On some limitations of data-driven weather forecasting models. arXiv e-prints, arXiv-2309.

Bonev, B., Kurth, T., Hundt, C., Pathak, J., Baust, M., Kashinath, K., & Anandkumar, A. (2023, July). Spherical fourier neural operators: Learning stable dynamics on the sphere. In International conference on machine learning (pp. 2806-2823). PMLR.

Bonev, B., Kurth, T., Mahesh, A., Bisson, M., Kossaifi, J., Kashinath, K., ... & Keller, A. (2025). FourCastNet 3: A geometric approach to probabilistic machine-learning weather forecasting at scale. arXiv preprint arXiv:2507.12144.

Bowler, N. E., Arribas, A., Mylne, K. R., Robertson, K. B., & Beare, S. E. (2008). The MOGREPS short-range ensemble prediction system. *Quarterly Journal of the Royal Meteorological Society: A journal of the atmospheric sciences, applied meteorology and physical oceanography*, *134*(632), 703-722.

Brenowitz, N. D., Cohen, Y., Pathak, J., Mahesh, A., Bonev, B., Kurth, T., ... & Pritchard, M. S. (2025). A practical probabilistic benchmark for ai weather models. Geophysical Research Letters, 52(7), e2024GL113656.

Camargo SJ, Wheeler MC, Sobel AH (2009) Diagnosis of the MJO modulation of tropical cyclogenesis using an empirical index. Journal of the Atmospheric Sciences 66, 3061–3074.

Chen, L., Zhong, X., Zhang, F., Cheng, Y., Xu, Y., Qi, Y., & Li, H. (2023). FuXi: A cascade machine learning forecasting system for 15-day global weather forecast. npj Climate and Atmospheric Science, 6(1), 190.

Cowan, T., Wheeler, M. C., & Marshall, A. G. (2023). The combined influence of the Madden–Julian Oscillation and El Niño–Southern Oscillation on Australian rainfall. Journal of Climate, 36(2), 313–334. https://doi.org/10.1175/JCLI-D-22-0357.1[1]

DeMaria, M., Franklin, J. L., Chirokova, G., Radford, J., DeMaria, R., Musgrave, K. D., & Ebert-Uphoff, I. (2025). An operations-based evaluation of tropical cyclone track and intensity forecasts from artificial intelligence weather prediction models. Artificial Intelligence for the Earth Systems, 4(4), 240085.

Diao, C., & Barnes, E. A. (2025). Assessing MJO Tropical-Extratropical Teleconnections in Deep Learning Weather Prediction Models. Authorea Preprints.

Ding, R., Li, J., & Seo, K. H. (2010). Predictability of the Madden–Julian oscillation estimated using observational data. Monthly Weather Review, 138(3), 1004-1013.

Dosovitskiy, A., Beyer, L., Kolesnikov, A., Weissenborn, D., Zhai, X., Unterthiner, T., ... & Houlsby, N. (2020). An image is worth 16x16 words. arXiv preprint arXiv:2010.11929, 7.

Dunstan, T., Strickson, O., Bennett, T., Bowyer, J., Burnand, M., Chappell, J., ... & Hosking, J. S. (2025). FastNet: Improving the physical consistency of machine-learning weather prediction models through loss function design. arXiv preprint arXiv:2509.17601.

Gillett NP, Kell TD, Jones PD (2006) Regional climate impacts of the southern annular mode. Geophysical Research Letters 33, L23704. https://doi.org/10.1029/2006GL027721.

Guan, H., Arcomano, T., Chattopadhyay, A., & Maulik, R. (2024). LUCIE: A Lightweight Uncoupled ClImate Emulator with long-term stability and physical consistency for O (1000)-member ensembles. arXiv preprint arXiv:2405.16297.

Hakim, G. J., & Masanam, S. (2024). Dynamical tests of a deep-learning weather prediction model. Artificial Intelligence for the Earth Systems.

Harrison, D. R., McGovern, A., Karstens, C. D., Bostrom, A., Demuth, J. L., Jirak, I. L., & Marsh, P. T. (2025). An assessment of how domain experts evaluate machine learning in operational meteorology. Weather and Forecasting, 40(3), 393-410.

Hendon HH, Wheeler MC, Zhang C (2007a). Seasonal dependence of the MJO-ENSO relationship. Journal of Climate 20(3), 531–543.

Hendon HH, Thompson DWJ, Wheeler MC (2007b), Australian rainfall and surface temperature variations associated with the Southern Hemisphere annular mode. Journal of Climate 20, 2452–2467.

Hersbach, H., Bell, B., Berrisford, P., Hirahara, S., Horányi, A., Muñoz-Sabater, J., ... & Thépaut, J. N. (2020). The ERA5 global reanalysis. Quarterly Journal of the Royal Meteorological Society, 146(730), 1999-2049.

Hudson, D., Alves, O., Hendon, H.H., et al. 2017: ACCESS-S1: The new Bureau of Meteorology multi-week to seasonal prediction system. Journal of Southern Hemisphere Earth Systems Science, 67:3 132-159 doi: 10.22499/3.6703.001

Husain, S. Z., Separovic, L., Caron, J. F., Aider, R., Buehner, M., Chamberland, S., ... & Zadra, A. (2025). Leveraging data-driven weather models for improving numerical weather prediction skill through large-scale spectral nudging. Weather and Forecasting, 40(9), 1749-1771.

Jiang, X., Adames, Á. F., Kim, D., Maloney, E. D., Lin, H., Kim, H., Zhang, C., DeMott, C. A., & Klingaman, N. P. (2020). Fifty years of research on the Madden-Julian Oscillation: Recent progress, challenges, and perspectives. Journal of Geophysical Research: Atmospheres, 125(17).

Kalnay, E., Kanamitsu, M., Kistler, R., Collins, W., Deaven, D., Gandin, L., ... & Joseph, D. (2018). The NCEP/NCAR 40-year reanalysis project. In Renewable energy (pp. Vol1_146-Vol1_194). Routledge.

Kovachki, N., Li, Z., Liu, B., Azizzadenesheli, K., Bhattacharya, K., Stuart, A., & Anandkumar, A. (2023). Neural operator: Learning maps between function spaces with applications to pdes. Journal of Machine Learning Research, 24(89), 1-97.

Lam, R., Sanchez-Gonzalez, A., Willson, M., Wirnsberger, P., Fortunato, M., Alet, F., ... & Battaglia, P. (2023). Learning skillful medium-range global weather forecasting. Science, 382(6677), 1416-1421.

Lang, S., Alexe, M., Chantry, M., Dramsch, J., Pinault, F., Raoult, B., ... & Rabier, F. (2024a). AIFS-ECMWF's data-driven forecasting system. arXiv preprint arXiv:2406.01465.

Lang, S., Alexe, M., Clare, M. C., Roberts, C., Adewoyin, R., Bouallègue, Z. B., ... & Leutbecher, M. (2024b). AIFS-CRPS: Ensemble forecasting using a model trained with a loss function based on the Continuous Ranked Probability Score. arXiv preprint arXiv:2412.15832.

Liu, X., Chen, J., Zhu, Y., Liu, Y., Chen, F., Huo, Z., ... & Gong, Y. (2025). Global Ensemble Weather Prediction from a Deep Learning–Based Model (Pangu-Weather) with the Initial Condition Perturbations of CMA-GEPS. Advances in Atmospheric Sciences, 1-25.

Mahesh, A., Collins, W., Bonev, B., Brenowitz, N., Cohen, Y., Elms, J., ... & Willard, J. (2024). Huge ensembles part i: Design of ensemble weather forecasts using spherical fourier neural operators. arXiv preprint arXiv:2408.03100.

Marshall, A. G., Hudson, D., Wheeler, M. C., Hendon, H. H., & Alves, O. (2012). Simulation and prediction of the Southern Annular Mode and its influence on Australian intra-seasonal climate in POAMA. Climate Dynamics, 38(11–12), 2483–2502. https://doi.org/10.1007/s00382-011-1140-z

Marshall, A. G., and Hendon, H. H. (2019). Multi-week prediction of the Madden–Julian oscillation with ACCESS-S1. Clim. Dyn. 52, 2513–2528

Morisseau, H., Zhu, H., Hudson, D. & de Burgh-Day, C. (February, 2025). Object-oriented verification of TC-Jasper rainfall forecasts: machine learning model versus physical models. Bureau Research Report No. 106, Bureau of Meteorology, Melbourne, Victoria https://nla.gov.au/nla.obj-3603384794

Neena, J. M., Lee, J. Y., Waliser, D., Wang, B., & Jiang, X. (2014). Predictability of the Madden–Julian Oscillation in the Intraseasonal Variability Hindcast Experiment (ISVHE). Journal of Climate, 27(12), 4531–4543. https://doi.org/10.1175/JCLI-D-13-00624.1[1]

Olivetti, L., & Messori, G. (2025). Whose weather is it? A fairness framework for data-driven weather forecasting. *Environmental Research Letters*, *20*(12), 121006.

Pathak, J., Subramanian, S., Harrington, P., Raja, S., Chattopadhyay, A., Mardani, M., ... & Anandkumar, A. (2022). Fourcastnet: A global data-driven high-resolution weather model using adaptive fourier neural operators. arXiv preprint arXiv:2202.11214.

Price, I., Sanchez-Gonzalez, A., Alet, F., Andersson, T. R., El-Kadi, A., Masters, D., ... & Willson, M. (2025). Probabilistic weather forecasting with machine learning. *Nature*, *637*(8044), 84-90.

Rasp, S., Hoyer, S., Merose, A., Langmore, I., Battaglia, P., Russell, T., ... & Sha, F. (2024). WeatherBench 2: A benchmark for the next generation of data-driven global weather models. Journal of Advances in Modeling Earth Systems, 16(6), e2023MS004019.

Robertson, A.W. and Vitart, F. (2019) Sub-seasonal to seasonal prediction. Elsevier. Available from: https://doi.org/10.1016/c2016-0-01594-2

Sahu, P. L., Sandeep, S., & Kodamana, H. (2025). Evaluating global machine learning models for tropical cyclone dynamics and thermodynamics. Journal of Geophysical Research: Machine Learning and Computation, 2(2), e2025JH000594.

Sun, Y. Q., Hassanzadeh, P., Zand, M., Chattopadhyay, A., Weare, J., & Abbot, D. S. (2025). Can AI weather models predict out-of-distribution gray swan tropical cyclones?. Proceedings of the National Academy of Sciences, 122(21), e2420914122.

Thompson DWJ, Wallace JM (2000) Annular Modes in the Extratropical Circulation. Part I: Month-to-Month Variability. Journal of Climate 13, 1000–1016.

Vitart, F. (2017). Madden–Julian Oscillation Prediction and Teleconnections in the S2S Database. Quarterly Journal of the Royal Meteorological Society, 143:2210–2220

Watt-Meyer, O., Dresdner, G., McGibbon, J., Clark, S. K., Henn, B., Duncan, J., ... & Bretherton, C. S. (2023). ACE: A fast, skillful learned global atmospheric model for climate prediction. arXiv preprint arXiv:2310.02074.

Wedd R et al (2022) ACCESS-S2: the upgraded Bureau of Meteorology multi-week to seasonal prediction system. Journal of Southern Hemisphere Earth Systems Science 72(3), 218–242. doi:10.1071/ES22026

Weyn, J. A., Durran, D. R., Caruana, R., & Cresswell-Clay, N. (2021). Sub-seasonal forecasting with a large ensemble of deep-learning weather prediction models. Journal of Advances in Modeling Earth Systems, 13(7), e2021MS002502.

Wheeler, M. C., & Hendon, H. H. (2004). An all-season real-time multivariate MJO index: Development of an index for monitoring and prediction. Monthly weather review, 132(8), 1917-1932.

Wheeler MC, McBride JL (2005) Australian-Indonesian monsoon. In: W.K.M. Lau and D.E. Waliser (eds), Intraseasonal Variability in the Atmosphere-Ocean Climate System. Praxis, Springer Berlin Heidelberg, pages 125-173.

Wheeler, M. C., and H. H. Hendon, S. Cleland, H. Meinke, and A. Donald, 2009: Impacts of the Madden–Julian oscillation on Australian rainfall and circulation. J. Climate, 22, 1482–1498, https://doi.org/10.1175/2008JCLI2595.1

White, C. J., Domeisen, D. I., Acharya, N., Adefisan, E. A., Anderson, M. L., Aura, S., ... & Wilson, R. G. (2022). Advances in the application and utility of subseasonal-to-seasonal predictions. Bulletin of the American Meteorological Society, 103(6), E1448-E1472.

Xiong, X., Hu, W., Zhou, S., Bi, K., Xie, L., Liu, Y., ... & Tian, Q. (2025). Bridging the Gap Between Bayesian Deep Learning and Ensemble Weather Forecasts. arXiv preprint arXiv:2511.14218.

Zhong, X., Chen, L., Li, H., Buizza, R., Liu, J., Feng, J., ... & Lu, B. (2025). FuXi-ENS: A machine learning model for efficient and accurate ensemble weather prediction. Science Advances, 11(44), eadu2854.